%% file: ms.tex
\documentclass[fleqn, usenatbib]{mnras}
\usepackage{style} 
\usepackage{booktabs,array} 

\title[PNG with DESI Imaging]{Local primordial non-Gaussianity from the large-scale clustering of photometric DESI luminous red galaxies}

\input{sections/authorlist}

\begin{document}
\label{firstpage}
\pagerange{\pageref{firstpage}--\pageref{lastpage}}
\maketitle
\clearpage
\input{sections/abstract}


\input{sections/introduction}
\input{sections/data}
\input{sections/methodology}
\input{sections/results}
\input{sections/conclusion}
\input{sections/acknowledgement}
\input{sections/dataavailability}

\bibliographystyle{mnras}
\bibliography{refs} 
\appendix
\include{sections/extra}

\bsp	
\label{lastpage}

\end{document}

%% file: sections/authorlist.tex
\author[M. Rezaie et al.]{Mehdi Rezaie$^{1,2}$, Ashley J. Ross$^{2}$, Hee-Jong Seo$^{3,4}$, Hui Kong$^{5}$, \newauthor Anna Porredon$^{6,7}$, Lado Samushia$^{1}$, Edmond Chaussidon$^{8}$, Alex Krolewski$^{9,10,11}$, \newauthor Arnaud de Mattia$^{8}$, Florian Beutler$^{6}$, Jessica Nicole Aguilar$^{4}$, Steven Ahlen$^{12}$, \newauthor Shadab Alam$^{13}$, Santiago Avila$^{5}$, Benedict Bahr-Kalus$^{14}$, Jose Bermejo-Climent$^{15}$, \newauthor David Brooks$^{16}$, Todd Claybaugh$^{4}$, Shaun Cole$^{17}$, Kyle Dawson$^{18}$, \newauthor Axel de la Macorra$^{19}$, Peter Doel$^{16}$, Andreu Font-Ribera$^{5}$, Jaime E. Forero-Romero$^{20,21}$, \newauthor Satya Gontcho A Gontcho$^{4}$, Julien Guy$^{4}$, Klaus Honscheid$^{2, 22}$, Dragan Huterer$^{23}$, \newauthor Theodore Kisner$^{4}$, Martin Landriau$^{4}$, Michael Levi$^{4}$, Marc Manera$^{5,24}$, Aaron Meisner$^{25}$, \newauthor Ramon Miquel$^{5, 26}$, Eva-Maria Mueller$^{27}$, Adam Myers$^{28}$, Jeffrey A. Newman$^{29}$, \newauthor Jundan Nie$^{30}$, Nathalie Palanque-Delabrouille$^{4,8}$, Will Percival$^{9,10,11}$, Claire Poppett$^{4,31}$, \newauthor Graziano Rossi$^{32}$, Eusebio Sanchez$^{33}$, Michael Schubnell$^{23}$, Gregory Tarlé$^{23}$, \newauthor Benjamin Alan Weaver$^{25}$, Christophe Yèche$^{8}$, Zhimin Zhou$^{30}$, Hu Zou$^{30}$
\\~\\
$^{1}$Department of Physics, Kansas State University, 116 Cardwell Hall, Manhattan, KS 66506, USA\\ 
$^{2}$Center for Cosmology and AstroParticle Physics, The Ohio State University, 191 West Woodruff Avenue, Columbus, OH 43210, USA\\ 
$^{3}$Department of Physics \& Astronomy, Ohio University, Athens, OH 45701, USA\\ $^{4}$Lawrence Berkeley National Laboratory, 1 Cyclotron Road, Berkeley, CA 94720, USA\\ 
$^{5}$Institut de F\'{i}sica d’Altes Energies (IFAE), The Barcelona Institute of Science and Technology, Campus UAB, 08193 Bellaterra Barcelona, Spain\\ 
$^{6}$Institute for Astronomy, University of Edinburgh, Royal Observatory, Blackford Hill, Edinburgh EH9 3HJ, UK\\ 
$^{7}$Ruhr University Bochum, Faculty of Physics and Astronomy, Astronomical Institute, German Centre for Cosmological Lensing, 44780 Bochum, Germany\\ 
$^{8}$IRFU, CEA, Universit\'{e} Paris-Saclay, F-91191 Gif-sur-Yvette, France\\ 
$^{9}$Waterloo Centre for Astrophysics, University of Waterloo, 200 University Ave W, Waterloo, ON N2L 3G1, Canada\\ 
$^{10}$Perimeter Institute for Theoretical Physics, 31 Caroline St. North, Waterloo, ON N2L 2Y5, Canada\\ 
$^{11}$Department of Physics and Astronomy, University of Waterloo, 200 University Ave W, Waterloo, ON N2L 3G1, Canada\\ 
$^{12}$Physics Department, Boston University, 590 Commonwealth Avenue, Boston, MA 02215, USA\\ 
$^{13}$Tata Institute of Fundamental Research, Homi Bhabha Road, Mumbai 400005, India\\ 
$^{14}$Korea Astronomy and Space Science Institute, 776, Daedeokdae-ro, Yuseong-gu, Daejeon 34055, Republic of Korea\\ 
$^{15}$Department of Physics \& Astronomy, University of Rochester, 206 Bausch and Lomb Hall, P.O. Box 270171, Rochester, NY 14627-0171, USA\\ 
$^{16}$Department of Physics \& Astronomy, University College London, Gower Street, London, WC1E 6BT, UK\\ 
$^{17}$Institute for Computational Cosmology, Department of Physics, Durham University, South Road, Durham DH1 3LE, UK\\ 
$^{18}$Department of Physics and Astronomy, The University of Utah, 115 South 1400 East, Salt Lake City, UT 84112, USA\\ 
$^{19}$Instituto de F\'{\i}sica, Universidad Nacional Aut\'{o}noma de M\'{e}xico,  Cd. de M\'{e}xico  C.P. 04510,  M\'{e}xico\\ 
$^{20}$Departamento de F\'isica, Universidad de los Andes, Cra. 1 No. 18A-10, Edificio Ip, CP 111711, Bogot\'a, Colombia\\ 
$^{21}$Observatorio Astron\'omico, Universidad de los Andes, Cra. 1 No. 18A-10, Edificio H, CP 111711 Bogot\'a, Colombia\\ 
$^{22}$Department of Physics, The Ohio State University, 191 West Woodruff Avenue, Columbus, OH 43210, USA\\ 
$^{23}$Department of Physics, University of Michigan, Ann Arbor, MI 48109, USA\\
$^{24}$Departament de F\'{i}sica, Universitat Aut\`{o}noma de Barcelona, 08193 Bellaterra (Barcelona), Spain\\ 
$^{25}$NSF's NOIRLab, 950 N. Cherry Ave., Tucson, AZ 85719, USA\\ 
$^{26}$Instituci\'{o} Catalana de Recerca i Estudis Avan\c{c}ats, Passeig de Llu\'{\i}s Companys, 23, 08010 Barcelona, Spain\\ 
$^{27}$Department of Physics and Astronomy, University of Sussex, Brighton BN1 9QH, U.K\\ 
$^{28}$Department of Physics \& Astronomy, University  of Wyoming, 1000 E. University, Dept.~3905, Laramie, WY 82071, USA\\ 
$^{29}$Department of Physics \& Astronomy, University of Pittsburgh, 3941 O'Hara Street, Pittsburgh, PA 15260, USA\\ 
$^{30}$National Astronomical Observatories, Chinese Academy of Sciences, A20 Datun Rd., Chaoyang District, Beijing, 100012, P.R. China\\ 
$^{31}$Space Sciences Laboratory, University of California, Berkeley, 7 Gauss Way, Berkeley, CA  94720, USA\\ 
$^{32}$Department of Physics and Astronomy, Sejong University, Seoul, 143-747, Korea\\ 
$^{33}$CIEMAT, Avenida Complutense 40, E-28040 Madrid, Spain}

%% file: sections/abstract.tex
\begin{abstract}
We use angular clustering of luminous red galaxies from the Dark Energy Spectroscopic Instrument (DESI) imaging surveys to constrain the local primordial non-Gaussianity parameter $\fnl$. Our sample comprises over 12 million targets, covering 14,000 square degrees of the sky, with redshifts in the range $0.2< z < 1.35$. We identify Galactic extinction, survey depth, and astronomical seeing as the primary sources of systematic error, and employ linear regression and artificial neural networks to alleviate non-cosmological excess clustering on large scales. Our methods are tested against simulations with and without $\fnl$ and systematics, showing superior performance of the neural network treatment. The neural network with a set of nine imaging property maps passes our systematic null test criteria, and is chosen as the fiducial treatment. Assuming the universality relation, we find $\fnl = 34^{+24(+50)}_{-44(-73)}$ at 68\%(95\%) confidence. We apply a series of robustness tests (e.g., cuts on imaging, declination, or scales used) that show consistency in the obtained constraints. We study how the regression method biases the measured angular power-spectrum and degrades the $\fnl$ constraining power. The use of the nine maps more than doubles the uncertainty compared to using only the three primary maps in the regression. Our results thus motivate the development of more efficient methods that avoid over-correction, protect large-scale clustering information, and preserve constraining power. Additionally, our results encourage further studies of $\fnl$ with DESI spectroscopic samples, where the inclusion of 3D clustering modes should help separate imaging systematics and lessen the degradation in the $\fnl$ uncertainty.
\end{abstract}

%% file: sections/introduction.tex
\section{Introduction}
\label{sec:introduction}
Inflation is a widely accepted paradigm in modern cosmology that explains many important characteristics of our Universe. It predicts that the early Universe underwent a period of accelerated expansion, resulting in the observed homogeneity and isotropy of the Universe on large scales \citep{PhysRevD.23.347, LINDE1982389,  PhysRevLett.48.1220}. After the period of inflation, the Universe entered a phase of reheating in which primordial perturbations were generated, setting the initial seeds for structure formation \citep{kofman1994reheating, bassett2006inflation, lyth2009primordial}. Although inflation is widely accepted as a compelling explanation, the characteristics of the field or fields that drove the inflationary expansion remain largely unknown in cosmology. While early studies of the cosmic microwave background (CMB) and large-scale structure (LSS) suggested that primordial fluctuations are both Gaussian and scale-invariant \citep{Komatsu_2003,PhysRevD.69.103501, guth2005inflationary}, some alternative classes of inflationary models predict different levels of non-Gaussianities in the primordial gravitational field. Non-Gaussianities are a measure of the degree to which the distribution of matter in the Universe deviates from a Gaussian distribution, which would have important implications for the growth of structure and galaxies in the Universe \citep[see, e.g.,][]{2010AdAst2010E..64V, 2010CQGra..27l4011D, Biagetti2019Galax...7...71B}.

In its simplest form, local primordial non-Gaussianity (PNG) is parameterized by the non-linear coupling constant $\fnl$\citep{komatsu2001acoustic}:
\begin{equation}
 \Phi = \phi + \fnl [\phi^{2} - <\phi^{2}>],
\end{equation}
where $\Phi$ is the primordial curvature perturbation and $\phi$ is assumed to be a Gaussian random field. Local-type PNG generates a primordial bispectrum, which peaks in the squeezed triangle configuration where one of the three wave vectors is much smaller than the other two. This means that one of the modes is on a much larger scale than the other two, and this mode couples with the other two modes to generate a non-Gaussian signal, which then affects the local number density of galaxies. The coupling between the short and long wavelengths induces a distinct bias in the galaxy distribution, which leads to a $k^{-2}$-dependent feature in the two-point clustering of galaxies and quasars \citep{dalal2008imprints}. Obtaining reliable, accurate, and robust constraints on $\fnl$ is crucial in advancing our understanding of the dynamics of the early Universe. For instance, the standard single-field slow-roll inflationary model predicts a small value of $\fnl \sim 0.01$ \citep[see, e.g.,][]{2003JHEP...05..013M}. On the other hand, some alternative inflationary scenarios involve multiple scalar fields that can interact with each other during inflation, leading to the generation of larger levels of non-Gaussianities. These models predict considerably larger values of $\fnl$ that can reach up to $100$ or higher \citep[see, e.g.,][for a review]{2010AdAst2010E..72C}. With $\sigma (\fnl)\sim 1$, we can rule out or confirm specific models of inflation and gain insight into the physics that drove the inflationary expansion \citep[see, e.g.,][]{alvarez2014arXiv1412.4671A, de2017next}.

The current tightest bound on $\fnl$ comes from Planck's bispectrum measurement of CMB anisotropies, $\fnl=0.9\pm 5.1$ \citep{akrami2019planck}. Limited by cosmic variance, CMB data cannot enhance the statistical precision of $\fnl$ measurements enough to break the degeneracy amongst various inflationary paradigms \citep[see, e.g.,][]{2016arXiv161002743A, ade2019simons}. On the other hand, LSS surveys probe a 3D map of the Universe, and thus provide more modes to limit $\fnl$. However, nonlinearities raised from structure formation pose a serious challenge for measuring $\fnl$ with the three-point clustering of galaxies, and these nonlinear effects are non-trivial to model and disentangle from the primordial signal \citep{baldauf2011galaxy, baldauf2011primordial}. Currently, the most precise constraints on $\fnl$ from LSS reach a level of $\sigma(\fnl) \sim 20-30$, with the majority of the constraining power coming from the two-point clustering statistics that utilize the scale-dependent bias effect \citep{slosar2008constraints,2013MNRAS.428.1116R,2019JCAP...09..010C, mueller2022primordial, 2022PhRvD.106d3506C, 2022arXiv220111518D}. Surveying large areas of the sky can unlock more modes and help improve these constraints. 

The Dark Energy Spectroscopic Instrument (DESI) is ideally suited to enable excellent constraints on primordial non-Gaussianity from the galaxy distribution. DESI uses $5000$ robotically-driven fibers to simultaneously collect spectra of extra-galactic objects \citep{2013arXiv1308.0847L, 2016arXiv161100037D, 2023AJ....165....9S}. DESI is designed to deliver an unparalleled volume of spectroscopic data covering $\sim 14,000$ square degrees that promises to deepen our understanding of the energy contents of the Universe, neutrino masses, and the nature of gravity \citep{2022AJ....164..207D}. Moreover, DESI alone is expected to improve our constraints on local PNG down to $\sigma(\fnl)=5$, assuming systematic uncertainties are under control \citep{aghamousa2016desi}. With multi-tracer techniques \citep{PhysRevLett.102.021302}, cosmic variance can be further reduced to allow surpassing CMB-like constraints \citep{2015ApJ...814..145A}. For instance, the distortion of CMB photons around foreground masses, which is referred to as CMB lensing, provides an additional probe of LSS, but from a different vantage point. We can significantly reduce statistical uncertainties below $\sigma(\fnl)\sim 1$ by cross-correlating LSS data with CMB-lensing, or other tracers of matter, such as 21 cm intensity mapping  \citep[see, e.g.,][]{schmittfull2018PhRvD, Heinrich2022AAS...24020203H, 2023arXiv230102406J, 2023arXiv230308901S}.  
 
However, further work is needed to fully harness the potential of the scale-dependent bias effect in constraining $\fnl$ with LSS. The amplitude of the $\fnl$ signal in the galaxy distribution is proportional to the bias parameter $b_{\phi}$, such that $\Delta b \propto b_{\phi}\fnl k^{-2}$. Assuming the universality relation, $b_{\phi} \sim (b - p)$, where $b$ is the linear halo bias and $p=1$ is a parameter that describes the response of galaxy formation to primordial potential perturbations in the presence of local PNG \citep[see, e.g.,][]{slosar2008constraints}. The value of $p$ is not very well constrained for other tracers of matter \citep{2020JCAP...12..013B, 2020JCAP...12..031B}, and \cite{2022JCAP...11..013B} showed that marginalizing over $p$ even with wide priors leads to biased $\fnl$ constraints because of parameter space projection effects. More simulation-based studies are necessary to investigate the halo-assembly bias and the relationship between $b_{\phi}$ and $b$ for various galaxy samples. For instance, \cite{2023JCAP...01..023L} used N-body simulations to investigate secondary halo properties, such as concentration, spin and sphericity of haloes, and found that halo spin and sphericity preserve the universality of the halo occupation function while halo concentration significantly alters the halo function.  Without better-informed priors on $p$, it is argued that the scale-dependent bias effect can only be used to constrain the $b_{\phi}\fnl$ term \citep[see, e.g.,][]{2020JCAP...12..031B}. However, regardless of the specific value of $p$, a nonzero detection of $b_{\phi}\fnl$ implies the presence of local PNG, given that $b_{\phi}$ is greater than zero. In this work, we assume the universality relation that links $b_{\phi}$ to $b-p$ and, further, fix the value of $p$.  

In addition to the theoretical uncertainties, measuring $\fnl$ through the scale-dependent bias effect is a difficult task due to various imaging systematic effects that can modulate the galaxy power spectrum on large scales. The imaging systematic effects often induce wide-angle variations in the density field, and in general, any large-scale variations can translate into an excess signal in the power spectrum \citep[see, e.g.,][]{huterer2013calibration}, that can be misinterpreted as the signature of non-zero local PNG \citep[see, e.g.,][]{PhysRevLett.106.241301}. Such spurious variations can be caused by Galactic foregrounds, such as dust extinction and stellar density, or varying imaging conditions, such as astrophysical seeing and survey depth \citep[see, e.g.,][]{ross2011}. The imaging systematic issues have made it challenging to accurately measure $\fnl$, as demonstrated in previous efforts to constrain it using the large-scale clustering of galaxies and quasars \citep[see, e.g.,][]{2013MNRAS.428.1116R,pullen2013systematic, Ho2015JCAP...05..040H}, and it is anticipated that they will be particularly problematic for wide-area galaxy surveys that observe regions of the night sky closer to the Galactic plane and that seek to incorporate more lenient selection criteria to accommodate fainter galaxies \citep[see, e.g,][]{kitanidis2020imaging}.

The primary objective of this paper is to utilize the scale-dependent bias signature in the angular power spectrum of galaxies selected from DESI imaging data to constrain the value of $\fnl$. With an emphasis on a careful treatment of imaging systematic effects, we aim to lay the groundwork for subsequent studies of local PNG with DESI spectroscopy. To prepare our sample for measuring such a subtle signal, we employ linear multivariate regression and artificial neural networks to mitigate spurious density fluctuations and ameliorate the excess clustering power caused by imaging systematics. We thoroughly investigate potential sources of systematic error, including survey depth, astronomical seeing, photometric calibration, Galactic extinction, and local stellar density. Our methods and results are validated against simulations, with and without imaging systematics.

This paper is structured as follows. Section \ref{sec:data} describes the galaxy sample from DESI imaging and lognormal simulations with, or without, PNG and synthetic systematic effects. Section \ref{sec:method} outlines the theoretical framework for modelling the angular power spectrum, strategies for handling various observational and theoretical systematic effects, and statistical techniques for measuring the significance of remaining systematics in our sample after mitigation. Our results are presented in Section \ref{sec:results}, and Section \ref{sec:conclusion} summarizes our conclusions and directions for future work.

%% file: sections/data.tex
\section{Data}
\label{sec:data}
Luminous red galaxies (LRGs) are massive galaxies that populate massive haloes, lack active star formation, and are highly biased tracers of the dark matter gravitational field \citep{postman1984ApJ...281...95P, kauffmann}. A distinct break around 4000 \AA~in the LRG spectrum is often utilized to determine their redshifts accurately. LRGs are widely targeted in previous galaxy redshift surveys \citep[see, e.g.,][]{eisenstein2001spectroscopic, prakash2016sdss}, and their clustering and redshift properties are well studied \citep[see, e.g.,][]{ross2020MNRAS.498.2354R, gilmarin2020MNRAS.498.2492G, bautista2021MNRAS.500..736B, chapman2022MNRAS.516..617C}. 

DESI is designed to collect spectra of millions of LRGs covering the redshift range $0.2<z<1.35$. DESI selects its targets for spectroscopy from the DESI Legacy Imaging Surveys, which consist of three ground-based surveys that provide photometry of the sky in the optical $g$, $r$, and $z$ bands. These surveys include the Mayall $z$-band Legacy Survey using the Mayall telescope at Kitt Peak \citep[MzLS;][]{dey2018overview}, the Beijing–Arizona Sky Survey using the Bok telescope at Kitt Peak \citep[BASS;][]{zou2017project}, and the Dark Energy Camera Legacy Survey on the Blanco 4m telescope \citep[DECaLS;][]{flaugher2015dark}. As shown in Figure \ref{fig:ng}, the BASS and MzLS programmes observed the same footprint in the North Galactic Cap (NGC) while the DECaLS programme observed both caps around the galactic plane; the BASS+MzLS footprint is separated from the DECaLS NGC at DEC $> 32.375$ degrees, although there is an overlap between the two regions for calibration purposes \citep{dey2018overview}. Additionally, the DECaLS programme integrates observations executed from the Blanco instrument under the Dark Energy Survey \citep{abbott2016dark}, which cover about $1130 \deg^{2}$ of the South Galactic Cap (SGC) footprint. The DESI imaging catalogues also integrate the $3.4$ (W1) and $4.6$ $\mu m$ (W2) infrared photometry from the Wide-Field Infrared Explorer \citep[WISE;][]{wise2010AJ....140.1868W, meisner2018RNAAS...2....1M}.  

\subsection{DESI imaging LRGs}
\begin{figure}
 \centering
 \includegraphics[width=0.5\textwidth]{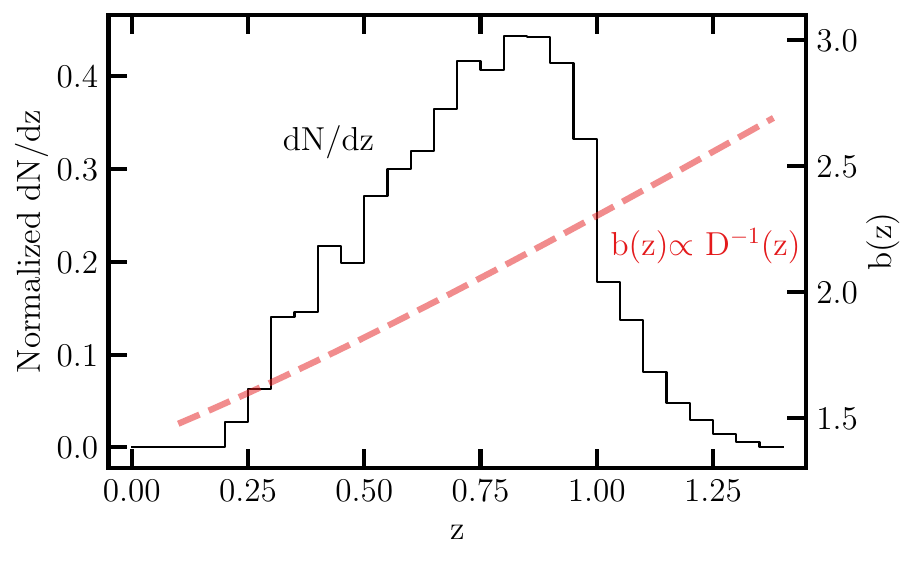}
 \caption{The redshift distribution (solid line and vertical scale on the left) and bias evolution (dashed line and vertical scale on the right) of the DESI LRG targets. The redshift distribution is determined from DESI spectroscopy \citep{desi2023sv}. The redshift evolution of the linear bias is supported by HOD fits to the angular clustering of the DESI LRG targets \citep{zhou2021clustering}, where $D(z)$ represents the growth factor.}
 \label{fig:nz}
\end{figure}

Our sample of LRGs is drawn from the DESI Legacy Imaging Surveys Data Release 9 \citep[DR9;][]{dey2018overview} using the color-magnitude selection criteria designed for the DESI 1\% survey \citep{desi2023sv}, described as the Survey Validation 3 (SV3) selection in more detail in \cite{zhou2022target}. The color-magnitude selection cuts are defined in the $g$, $r$, $z$ bands in the optical and $W1$ band in the infrared, as summarized in Table \ref{tab:ts}. The selection cuts vary for each imaging survey, but they are designed to achieve a nearly consistent density of approximately $800$ galaxies per square degree across a total area of roughly $14,000$ square degrees. Table \ref{tab:imaging} summarizes the mean galaxy density and area for each region. This is accomplished despite variations in survey efficiency and photometric calibration between DECaLS and BASS+MzLS. The implementation of these selection cuts in the DESI data processing pipeline is explained in \cite{myers2022}. The redshift distribution of our galaxy sample are inferred respectively from DESI spectroscopy during the Survey Validation phase \citep{desi2023sv}, and is shown via the solid curve in Figure \ref{fig:nz}. \cite{zhou2021clustering} analyzed the DESI LRG targets and found that the redshift evolution of the linear bias for these targets is consistent with a constant clustering amplitude and varies via $1/D(z)$, where $D(z)$ is the growth factor (as illustrated by the dashed red line in Figure \ref{fig:nz}). 

\begin{table*}
\caption{Color-magnitude selection criteria for the DESI LRG targets \citep{zhou2022target}. Magnitudes are corrected for Galactic extinction. The z-band fiber magnitude, $z_{\rm fiber}$, corresponds to the expected flux within a DESI fiber.}\label{tab:ts}
 \centerline{%
 \begin{tabular}{lll}
 \hline
 \hline
 \textbf{Footprint} & \textbf{Criterion} &\textbf{Description}\\
 \hline
 \hline  
 & $z_{\rm fiber} < 21.7$ & Faint limit \\
  DECaLS & $z - W1 > 0.8 \times (r - z) - 0.6$ & Stellar rejection \\
 & $[(g-r >1.3)~{\rm AND}~((g-r) > -1.55*(r-W1) + 3.13)]~{\rm OR}~(r -W 1 > 1.8)$ & Remove low-z galaxies \\
 & $[(r-W1 > (W1 - 17.26)*1.8)~{\rm AND}~(r - W1 > W1 - 16.36)]~{\rm OR}~(r-W1 > 3.29)$ & Luminosity cut \\ 
 \hline
 & $z_{\rm fiber} < 21.71$ & Faint limit \\
 BASS+MzLS & $z - W1 > 0.8 \times (r - z) - 0.6$ & Stellar rejection \\
 & $[(g-r >1.34)~{\rm AND}~((g-r) > -1.55*(r-W1) + 3.23)]~{\rm OR}~(r -W 1 > 1.8)$ & Remove low-z galaxies \\
 & $[(r-W1 > (W1 - 17.24)*1.83)~{\rm AND}~(r - W1 > W1 - 16.33)]~{\rm OR}~(r-W1 > 3.39)$ & Luminosity cut \\ 
 \hline
 \end{tabular}}
\end{table*}

\begin{figure*}
 \centering
 \includegraphics[width=\textwidth]{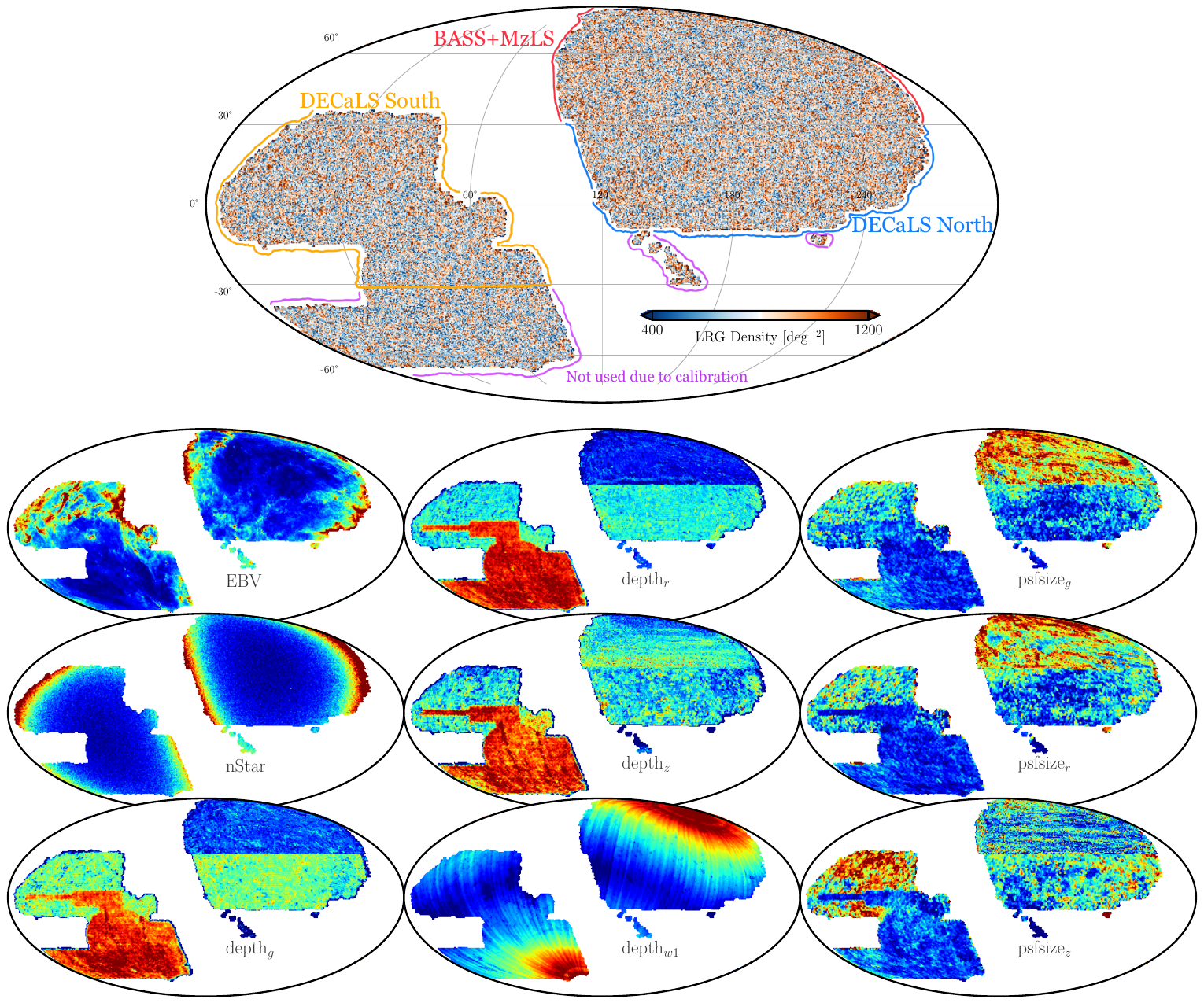}
 \caption{Top: The DESI LRG target density map before correcting for imaging systematic effects in Mollweide projection. The disconnected islands from the North footprint and parts of the South footprint with declination below $-30$ are removed from the sample for the analysis due to potential calibration issues (see text). Bottom: Mollweide projections of the imaging systematic maps (survey depth, astronomical seeing/psfsize, Galactic extinction, and local stellar density) in celestial coordinates. Not shown here are two external maps for the neutral hydrogen column density and photometric calibration, which are only employed for the robustness tests. The imaging systematic maps are colour-coded to show increasing values from blue to red.}
 \label{fig:ng}
\end{figure*}

The LRG sample is masked rigorously for foreground bright stars, bright galaxies, and clusters of galaxies\footnote{See \url{https://www.legacysurvey.org/dr9/bitmasks/} for maskbit definitions.} to further reduce stellar contamination \citep{zhou2022target}. Then, the sample is binned into \textsc{HEALPix} \citep{gorski2005healpix} pixels at $\textsc{nside}=256$, corresponding to pixels of about $0.25$ degrees on a side, to construct the 2D density map (as shown in the top panel of Figure \ref{fig:ng}). The LRG density is corrected for the pixel incompleteness and lost areas using a catalogue of random points, hereafter referred to as randoms, uniformly scattered over the footprint with the same cuts and masks applied. Moreover, the density of galaxies is matched to the randoms separately for each of the three data sections (BASS+MzLS, DECaLS North / South) so the mean density differences are mitigated (see Table \ref{tab:imaging}). The DESI LRG targets are selected brighter than the imaging survey depth limits, e.g., $g=24.4,~r=23.8,~{\rm and}~z=22.9$ for the median $5\sigma$ detection in AB mag in the DECaLS North region (Table \ref{tab:imaging}); and thus the LRG density map does not exhibit severe spurious fluctuations.

\subsubsection{Imaging systematic maps}
The effects of observational systematics in the DESI targets have been studied in great detail \cite[see, e.g.,][]{kitanidis2020imaging, zhou2021clustering, chaussidon2022angular}.  \cite{zhou2022target} has previously identified nine astrophysical properties as potential sources of imaging systematic errors in the DESI LRG targets. These imaging properties are mapped into \textsc{HEALPix} of \textsc{nside}$=256$. As illustrated by the $3\times3$ grid in the bottom panel of Figure \ref{fig:ng}, the maps include local stellar density constructed from point-like sources with a G-band magnitude in the range $12 \leq G < 17$ from the \textit{Gaia} DR2 \citep[see,][]{gaiadr2, myers2022}; Galactic extinction E[B-V] from \cite{schlegel1998maps}; survey depth (galaxy depth in $g$, $r$, and $z$ and PSF depth in W1) and astronomical seeing (i.e., point spread function, or psfsize) in $g$, $r$, and $z$. The depth maps have been corrected for extinction using the coefficients adapted from \cite{2011ApJ...737..103S}. Table \ref{tab:imaging} summarizes the median values for the imaging properties in each region. In addition to these nine maps, we consider two external maps for the neutral hydrogen column density (HI) from \cite{2016A&A...594A.116H} and photometric calibration in the z-band (CALIBZ) from \cite{desi2023sv} to further test the robustness of our analysis against unknown systematics.

\begin{table*}
\caption{Statistics for DESI imaging data. Median depths are for galaxy/point sources detected at $5\sigma$. Median psfsize values are computed with a depth-weighted average at each location on the sky.}
\begin{center}
\begin{tabular}{lccc}
\hline
\hline
    & BASS+MzLS & DECaLS North &DECaLS South \\
\hline
\hline
Mean galaxy density [deg$^{-2}$]     & 804  & 808  & 796 \\
Area [deg$^2$]                       & 4525 & 5257 & 5188 \\
Median extinction [mag]              & 0.02 & 0.03 & 0.05\\
Median stellar density [deg$^{-2}$]  & 667  & 629  & 629\\
Median $g$ galaxy depth [mag]        & 24.0 & 24.4 & 24.5 \\
Median $r$ galaxy depth [mag]        & 23.4 & 23.8 & 23.9\\
Median $z$ galaxy depth [mag]        & 23.0 & 22.9 & 23.1\\
Median $W1$ psf depth [mag]          & 21.6 & 21.4 & 21.4\\
Median $g$ psfsize [arcsec]          & 1.9  & 1.5  & 1.5\\
Median $r$ psfsize [arcsec]          & 1.7  & 1.4  & 1.3\\
Median $z$ psfsize [arcsec]          & 1.2  & 1.3  & 1.3\\
\hline
\end{tabular}
\end{center}
\label{tab:imaging}
\end{table*} 

The fluctuations in each imaging map are unique and tend to be correlated with the LRG density map. For instance, large-scale LRG density fluctuations could be caused by stellar density, extinction, or survey depth; while small scale-fluctuations could be caused by psfsize variations. Some regions of the DR9 footprint are removed from our analysis to avoid potential photometric calibration issues. These regions are either disconnected from the main footprint (e.g., the islands in the NGC with DEC $<-10$) or calibrated using different catalogues of standard stars (e.g., DEC $<-30$ in the SGC). The potential impact of not imposing these declination cuts on the LRG sample and our $\fnl$ constraints is explored in Section \ref{sec:results}. 

\begin{figure}
\centering
 \includegraphics[width=0.45\textwidth]{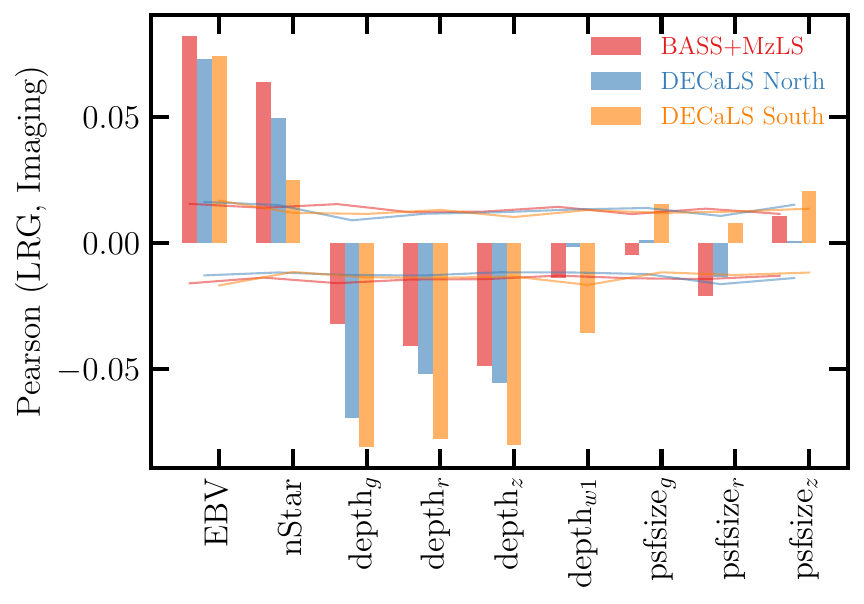} 
 \includegraphics[width=0.45\textwidth]{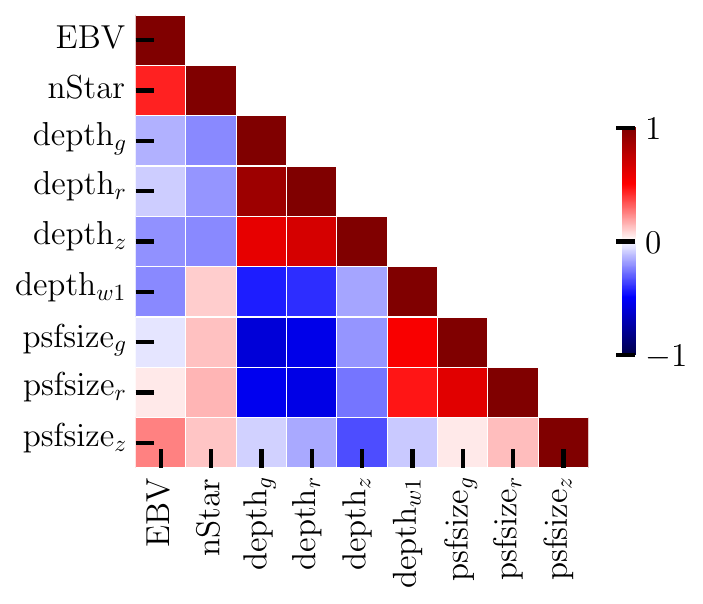}  
 \caption{Top: The Pearson correlation coefficient between the DESI LRG target density and imaging properties in BASS+MzLS, DECaLS North, and DECaLS South. Solid horizontal curves represent the $95\%$ confidence intervals estimated from simulations of lognormal density fields with $\fnl=0$. Bottom: The Pearson correlation matrix of imaging properties for the DESI footprint.}
 \label{fig:pcc}
\end{figure}

We employ the Pearson correlation coefficient to characterize the correlation between the galaxy density and imaging properties, which for two random variables $x$ and $y$ is given by, 
\begin{equation}
	\text{Pearson}~(x, y) = \frac{\sum (x_{i}-\bar{x})(y_{i}-\bar{y})}{\sqrt{\sum (x_{i}-\bar{x})^{2}\sum (y_{i}-\bar{y})^{2}}},
\end{equation}
where $\bar{x}$ and $\bar{y}$ represent the mean estimates of the random variables. Figure \ref{fig:pcc} shows the Pearson correlation coefficient between the DESI LRG target density map and the imaging systematics maps for the three imaging regions (DECaLS North, DECaLS South, and BASS+MzLS) in the top panel. The horizontal curves represent the $95\%$ confidence regions for no correlation and are constructed by cross-correlating 100 synthetic lognormal density fields, generated with $\fnl=0$, and the imaging systematic maps. Consistent among the different regions, there are statistically significant correlations between the LRG density and depth, extinction, and stellar density. There are less significant correlations between the LRG density and the $W1$-band depth and psfsize. The signs of the correlations imply that there are more targets where extinction is high, and less targets where depth is high. Another interpretation might be that more contaminants are targeted where depth is shallow. Figure \ref{fig:pcc} (bottom panel) shows the correlation matrix among the imaging systematic maps for the entire DESI footprint. Significant inner correlations exist among the imaging systematic maps themselves, especially between local stellar density and Galactic extinction; also, the $r$-band and $g$-band survey properties are more correlated with each other than with the $z$-band counterpart. Additionally, we compute the Spearman correlation coefficients between the LRG density and imaging systematic maps to assess whether or not the correlations are impacted by outliers in the imaging data, but find no substantial differences from Pearson.

\subsubsection{Treatment of imaging systematics}
There are several approaches for handling imaging systematic errors, broadly classified into data-driven and simulation-based modeling approaches \citep[see e.g.][]{ross2011, ashley2012MNRAS,Ross17,2012ApJ...761...14H,suchyta2016,delubac2016sdss, prakash2016sdss, Raichoor2017MNRAS.471.3955R, laurent2017clustering, Elvin18, 2018ApJ...863..110B, 2020MNRAS.495.1613R, kong2020,rezaie2021primordial,Everett_2022, chaussidon2022angular,Eggert_2023}. The general idea behind these approaches is to use the available data or simulations to learn or forward model the relationship between the observed target density and the imaging systematic maps, and to use this relationship, which is often described by a set of \textit{imaging weights}, to mitigate spurious fluctuations in the observed target density. Another techniques for reducing the effect of imaging systematics rely on cross-correlating different tracers of dark matter to ameliorate excess clustering signals, as each tracer might respond differently to a source of systematic error \citep[see, e.g.,][]{giannantonio2014improved}. These methods have their limitations and strengths \citep[see, e.g.,][for a review]{2021MNRAS.503.5061W}. In this paper, data-driven approaches, including linear multivariate regression and artificial neural networks, are applied to the data to correct for imaging systematic effects.

\textbf{Linear multivariate model}: The linear multivariate model only uses the imaging systematic maps up to the linear power to predict the number counts of the DESI LRG targets in pixel $i$,
\begin{equation}\label{eq:npred}
    N_{i} = \log ( 1 + \exp[\textbf{a}\cdot\textbf{x}_{i}+a_{0}]),
\end{equation}
where $a_{0}$ is a global offset, and $\textbf{a}\cdot\textbf{x}_{i}$ represents the inner product between the parameters, $\textbf{a}$, and the values for imaging systematics in pixel $i$, $\textbf{x}_{i}$. The Softplus functional form for $N_{i}$ is adapted to force the predicted galaxy counts to be positive \citep{dugas2001incorporating}. Then, Markov Chain Monte Carlo (MCMC) search is performed using the \textsc{emcee} package \citep{2013PASP..125..306F} to explore the parameter space by minimizing the negative Poisson log-likelihood between the actual and predicted number counts of galaxies.

Spatial coordinates are not included in $\textbf{x}_{i}$ to help avoid over-correction. As a result, the predicted number counts solely reflect the spurious density fluctuations that arise from varying imaging conditions. The number of pixels is substantially larger than the number of parameters for the linear model, and thus no training-validation-testing split is applied to the data for training the linear model. This aligns with the methodology used for training linear models in previous analyses \citep[see, e.g.,][]{zhou2022target}. The predicted galaxy counts are evaluated for each region using the marginalized mean estimates of the parameters, combined with those from other regions to cover the DESI footprint. The linear-based imaging weights are then defined as the inverse of the predicted target density, normalized to a median of unity.

\textbf{Neural network model}: Our neural network-based mitigation approach uses the implementation of fully connected feedforward neural networks from \cite{rezaie2021primordial}. With the neural network approach, $\textbf{a}\cdot\textbf{x}_{i}$ in Equation \ref{eq:npred} is replaced with $NN(\textbf{x}_{i}|\textbf{a})$, where $NN$ represents the fully connected neural network and $\textbf{a}$ denotes its parameters. The implementation, training, validation, and application of neural networks on galaxy survey data are presented in \cite{rezaie2021primordial}. We briefly summarize the methodology here. 

A fully connected feedforward neural network (also called a \textit{multi-layer perceptron}) is a type of artificial neural network where the neurons are arranged in layers, and each neuron in one layer is connected to every neuron in the next layer. The imaging systematic information flows only in one direction, from input to output. Each neuron applies a non-linear activation function (i.e., transformation) to the weighted sum of its inputs, which are the outputs of the neurons in the previous layer. The output of the last layer is the model prediction for the number counts of galaxies. Our architecture consists of three hidden layers with 20 rectifier activation functions on each layer, and a single neuron in the output layer. The rectifier is defined as ${\rm max}(0, x)$ to introduce nonlinearities in the neural network \citep{nair2010rectified}. This simple form of nonlinearity is very effective in enabling deep neural networks to learn more complex, non-linear relationships between the input imaging maps and output galaxy counts.

Compared with linear regression, neural networks potentially are more prone to over-fitting, i.e., excellent performance on training data and poor performance on validation (or test) data. Therefore, our analysis uses a training-validation-testing split to avoid over-fitting and ensure that the neural network is well-optimized. Specifically, $60\%$ of the LRG data is used for training, $20\%$ is used for validation, and $20\%$ is used for testing. The split is performed randomly aside from the locations of the pixels. We also test a geometrical split in which neighboring pixels belong to the same set of training, testing, or validation, but no significant performance difference is observed.

The neural networks are trained for up to 70 training epochs with the gradient descent \textsc{Adam} optimizer \citep{2017arXiv171105101L}, which iteratively updates the neural network parameters following the gradient of the negative Poisson log-likelihood. The step size of the parameter updates is controlled via the learning rate hyper-parameter, which is initialized with a grid search and is designed to dynamically vary between two boundary values of $0.001$ and $0.1$ to avoid local minima \citep[see again,][]{2016arXiv160803983L}. At each training epoch, the neural network model is applied to the validation set, and ultimately the model with the best performance on validation is identified and applied to the test set. The neural network models are tested on the entirety of the LRG sample with the technique of permuting the choice of the training, validation, or testing sets \citep{arlot2010survey}. With the cross-validation technique, the model predictions from the different test sets are aggregated together to form the predicted target density map into the DESI footprint. To reduce the error in the predicted number counts, we train an ensemble of 20 neural network models and average over the predictions. The imaging weights are then defined as the inverse of the predicted target density, normalized to a median of unity.

\subsection{Synthetic lognormal density fields}\label{ssec:mocks}
Density fluctuations of galaxies on large scales can be approximated with lognormal distributions \citep{coles1991, 2017MNRAS.466.1444C}. Unlike N-body simulations, simulating lognormal density fields is not computationally intensive, and allows quick and robust validation of data analysis pipelines. Lognormal simulations are therefore considered efficient for our study since the signature of local PNG appears on large-scales and small-scale clustering is not used in our analysis. The package \textsc{FLASK} \citep[Full-sky Lognormal Astro-fields Simulation Kit;][]{Xavier_2016} is employed to generate ensembles of synthetic lognormal density maps that mimic the bias, redshift, and angular distributions of the DESI LRG targets, as illustrated in Figure \ref{fig:nz} and \ref{fig:ng}. Two universes with $\fnl=0$ and $76.9$ are considered. A set of 1000 realizations is produced for every $\fnl$. The mocks are designed to match the clustering signal of the DESI LRG targets on scales insensitive to $\fnl$. The analysis adapts the fiducial BOSS cosmology \citep{2017MNRAS.470.2617A} which assumes a flat $\Lambda$CDM universe, including one massive neutrino with $m_{\nu}=0.06$ eV, Hubble constant $h = 0.68$, matter density $\Omega_{M}=0.31$, baryon density $\Omega_{b}=0.05$, and spectral index $n_{s}=0.967$. The amplitude of the matter density fluctuations on a scale of $8 h^{-1} \text{Mpc}$ is set as $\sigma_{8}=0.8225$. The same fiducial cosmology is used throughout this paper unless specified otherwise. Our robustness tests show that the none of the cosmological parameters can produce a $\fnl$-like signatures, and therefore, our analysis is not sensitive to the choice of fiducial cosmology.

\subsubsection{Contaminated mocks}
We employ the linear multivariate model (Equation \ref{eq:npred}) to introduce synthetic spurious fluctuations in the lognormal density fields, and validate our imaging systematic mitigation methods. The motivation for choosing a linear contamination model is to assess how much of the clustering signal can be removed by applying more flexible models, based on neural networks, for correcting less severe imaging systematic effects. The imaging systematic maps considered for the contamination model are extinction, depth in z, and psfsize in r. As shown in the Pearson correlation (Figure \ref{fig:pcc}) and will be discussed later in Section \ref{sec:systests}, the DESI LRG targets correlate strongly with these three maps. We fit for the parameters of the linear models with the MCMC process, executed separately on each imaging survey (BASS+MzLS, DECaLS North, and DECaLS South). Then, the imaging selection function for contaminating each simulation is uniquely determined by randomly drawing from the parameter space probed by MCMC, and then the results from each imaging survey are combined to form the DESI footprint. The clean density is then multiplied by the contamination model to induce systematics.  The same contamination model is used for both the $\fnl=0$ and $76.9$ simulations.

Similar to the imaging systematic treatment analysis for the DESI LRG targets, the neural network methods with various combinations of the imaging systematic maps are applied to each simulation, with and without PNG, and with and without systematics, to derive the imaging weights. Section \ref{sec:method} presents how the simulation results are incorporated to calibrate $\fnl$ biases due to over-correction. We briefly summarize two statistical tests based on the mean galaxy density contrast and the cross power spectrum between the galaxy density and the imaging systematic maps to assess the quality of the data and the significance of the remaining systematic effects (see, also, Rezaie et al. 2021). We calculate these statistics and compare the values to those measured from the clean mocks before looking at the auto power spectrum of the DESI LRG targets.

%% file: sections/methodology.tex
\section{Analysis techniques}
\label{sec:method} 
We address imaging systematics in DESI data by performing a separate treatment for each imaging region (e.g., DECaLS North) within the DESI footprint to reduce the impact of systematic effects specific to that region. Once the imaging systematic weights are obtained for each imaging region separately, we combine the data from all regions to compute the power spectrum for the entire DESI footprint to increase the overall statistical power and enable more robust measurements of $\fnl$. We then conduct robustness tests on the combined data to assess the significance of any remaining systematic effects.

\subsection{Power spectrum estimator}
We first construct the density contrast field from the LRG density, $\rho$,
\begin{align}\label{eq:delta}
    \delta_{g} &= \frac{\rho- \overline{\rho}}{\overline{\rho}},
\end{align}
where the mean galaxy density $\overline{\rho}$ is estimated from the entire LRG sample. As a robustness test, we also analyze the power spectrum from each imaging region individually, in which $\overline{\rho}$ is calculated separately for each region. Then, we use the pseudo angular power spectrum estimator \citep{hivon2002master},
\begin{equation}\label{eq:pusedocell}
        \tilde{C}_{\ell} = \frac{1}{2\ell +1} \sum_{m=-\ell}^{\ell} |a_{\ell m}|^{2},
\end{equation}
where the coefficients $a_{\ell m}$ are obtained by decomposing $\delta_{g}$ into spherical harmonics, $Y_{\ell m}$,
\begin{equation}\label{eq:alm}
        a_{\ell m} = \int d\Omega ~ \delta_{g} W Y^{*}_{\ell m},
\end{equation}
where $W$ represents the survey window that is described by the number of randoms normalized to the expected value.

We use the implementation of \texttt{anafast} from the \textsc{HEALPix} package \citep{gorski2005healpix} to do fast harmonic transforms (Equation \ref{eq:alm}) and estimate the pseudo angular power spectrum of the LRG targets and the cross power spectrum between the LRG targets and the imaging systematic maps.

\subsection{Modelling}
The estimator in Equation \ref{eq:pusedocell} yields a biased power spectrum when the survey sky coverage is incomplete. Specifically, the survey mask causes correlations between different harmonic modes \citep{beutler2014clustering,wilson2017rapid}, and the measured clustering power is smoothed on scales near the survey size. An additional potential cause of systematic error arises from the fact that the mean galaxy density used to construct the density contrast field (Equation \ref{eq:delta}) is estimated from the available data, rather than being known a priori. This introduces what is known as an integral constraint effect, which can cause the power spectrum on modes near the size of the survey to be artificially suppressed, effectively pushing it towards zero \citep{peacock1991large,de2019integral}. Since $\fnl$ is highly sensitive to the clustering power on these scales, it is crucial to account for these systematic effects in the model galaxy power spectrum to obtain unbiased $\fnl$ constraints \citep[see, also,][]{riquelme2022primordial}, which we describe below.
  
The other theoretical systematic issues are however subdominant in the angular power spectrum. For instance, relativistic effects generate PNG-like scale-dependent signatures on large scales, which interfere with measuring $\fnl$ with the scale-dependent bias effect using higher order multipoles of the 3D power spectrum \citep{wang2020}. Similarly, matter density fluctuations with wavelengths larger than survey size, known as super-sample modes, modulate the galaxy 3D power spectrum \citep{castorina2020JCAP}. In a similar way, the peculiar motion of the observer can mimic a PNG-like scale-dependent signature through aberration, magnification and the Kaiser-Rocket effect, i.e., a systematic dipolar apparent blue-shifting in the direction of the observer's peculiar motion \citep{2021JCAP...11..027B}.
  
\subsubsection{Angular power spectrum} 
The relationship between the linear matter power spectrum $P(k)$ and the projected angular power spectrum of galaxies is expressed by the following equation:
\begin{equation}\label{eq:cell}
C_{\ell} = \frac{2}{\pi}\int_{0}^{\infty}\frac{dk}{k}k^{3}P(k)|\Delta_{\ell}(k)|^{2} + N_{\rm shot},
\end{equation}
where $N_{\rm shot}$ is a scale-independent shot noise term. The projection kernel $\Delta_{\ell}(k) = \Delta^{\rm g}_{\ell}(k) + \Delta^{\rm RSD}_{\ell}(k) + \Delta^{\mu}_{\ell}(k)$ includes redshift space distortions and magnification bias, and determines the contribution of each wavenumber $k$ to the galaxy power spectrum on mode $\ell$. For more details on this estimator, refer to \cite{Padmanabhan2007}. The non-linearities in the matter power spectrum are negligible for the scales of interest \citep[see, e.g.,][]{Ho2015JCAP...05..040H}. For $\ell=40$, $\Delta_{\ell}(k)$ peaks at $k\sim 0.02~ h\text{Mpc}^{-1}$, which is above the non-linear regime. The FFTLog algorithm and its extension\footnote{\href{https://github.com/xfangcosmo/FFTLog-and-beyond}{github.com/xfangcosmo/FFTLog-and-beyond}} as implemented in \cite{fang2020beyond} are employed to calculate the integrals for the projection kernel $\Delta_{\ell}(k)$, which includes the $l^{\rm th}$ order spherical Bessel functions, $ j_{\ell}(kr)$, and its second derivatives,
\begin{align}
    \Delta^{\rm g}_{\ell}(k) &= \int \frac{dr}{r} r (b+\Delta b) D(r) \frac{dN}{dr} j_{\ell}(kr),\\
    \Delta^{\rm RSD}_{\ell}(k) &= - \int \frac{dr}{r} r f(r) D(r) \frac{dN}{dr} j^{\prime\prime}_{\ell}(kr),\\
    \Delta^{\mu}_{\ell}(k) &= - \ell(\ell+1) \int dr D(r) W_{\mu}(z) j_{\ell}(kr),
\end{align}
where $b$ is the linear bias (dashed curve in Figure \ref{fig:nz}), $D$ represents the linear growth factor normalized as $D(z=0)=1$, $f(r)$ is the growth rate, and $dN/dr$ is the redshift distribution of galaxies normalized to unity and described in terms of comoving distance\footnote{$dN/dr = (dN/dz)(dz/dr) \propto (dN/dz)H(z)$} (solid curve in Figure \ref{fig:nz}). The magnification bias window function $W_{\mu}(z)$ is
\begin{equation}
W_{\mu}(z) = (5s-2)\frac{3H^{2}_{0}\Omega_{m}(1+z)}{2c^{2}k^{2}} \int_{z}^{\infty} dz^{\prime}\frac{dN}{dz} \frac{r(z^{\prime}) - r(z)}{r(z^{\prime})r(z)},
\end{equation}
where $\Omega_{m}$ is the matter density, $H_{0}$ is the Hubble constant\footnote{$H_{0}=100~({\rm km}~{\rm s}^{-1})/(h^{-1}{\rm Mpc})$ and $k$ is in unit of $h {\rm Mpc}^{-1}$}, $c$ is the speed of light, and $s$ represents the slope of the number count function, a metric quantifying the response of the number density of galaxies to achromatic changes in brightness \citep{2008PhRvD..77b3512L}. The estimation of $s$ involves shifting all magnitudes by an infinitesimal amount and re-running the color-magnitude selection. \cite{zhou2023desi} developed a strategy to estimate $s$ for a fiber flux-selected sample like the DESI LRG targets, for which the impact of magnification on fiber flux depends on the shape parameters for each morphology type. Following the same strategy, the parameter $s$ is re-calculated for our selection of the DESI LRGs (DESI SV3)\footnote{Private communication with Dr. Rongpu Zhou.}: $s=0.951\pm 0.011$ for BASS+MzLS, $s=0.943 \pm 0.007$ for DECaLS North+DECaLS South, and $s=0.945\pm 0.006$ for DESI. For consistency, we fix $s$ to the above central values in our analysis.

The PNG-induced scale-dependent shift is given by \citep{slosar2008constraints}
\begin{equation}
\Delta b = b_{\phi}(z) \fnl \frac{3 \Omega_{m} H^{2}_{0}}{2 k^{2}T(k)D(z) c^{2}} \frac{g(\infty)}{g(0)},
\label{eq:scaledepbias}
\end{equation}
where $T(k)$ is the transfer function, and $g(\infty)/g(0) \sim 1.3$ with $g(z)\equiv (1+z) D(z)$ is the growth suppression due to non-zero $\Lambda$ because of our normalization of $D$ \citep[see, e.g.,][]{2010JCAP...07..013R, 2019MNRAS.485.4160M}. We assume the universality relation which directly relates $b_\phi$ to $b$ via $b_{\phi} = 2 \delta_{c}(b - p)$ with $\delta_{c}= 1.686$ representing the critical density for spherical collapse \citep{fillmore1984self}. We fix $p=1$ in our analysis and marginalize over b \citep[see, also,][]{slosar2008constraints,2010JCAP...07..013R,2013MNRAS.428.1116R}.

\subsubsection{Survey geometry and integral constraint}
\begin{figure}
    \centering
    \includegraphics[width=0.45\textwidth]{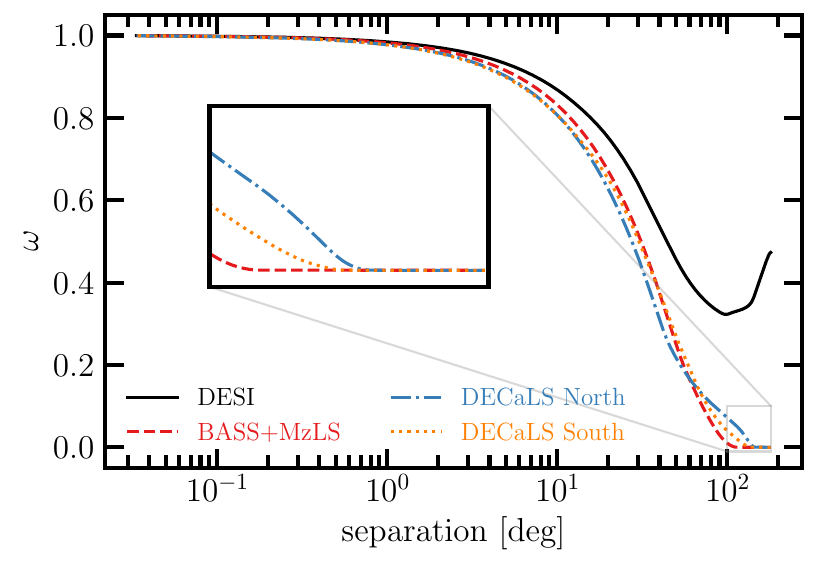}
    \caption{The survey mask correlation functions across the imaging regions forming the DESI footprint, plotted against angular separation. The inset focuses on correlations within the angular range of $100$ to $180$ degrees.}
    \label{fig:mask2pf}
\end{figure}

\begin{figure}
\centering
\includegraphics[width=0.45\textwidth]{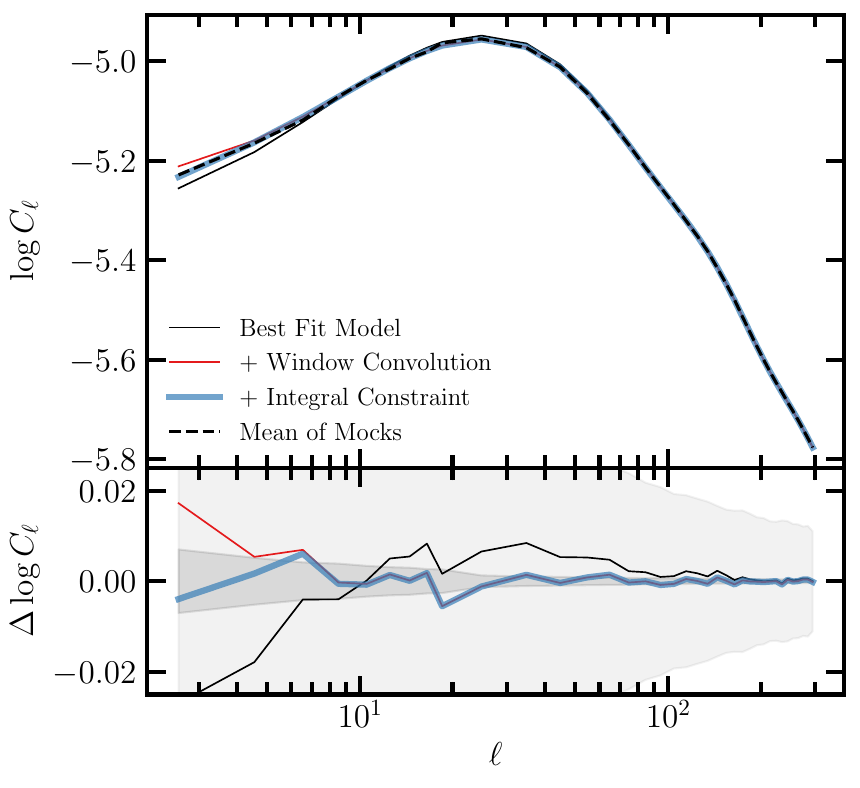}
\caption{The mean power spectrum from the $\fnl=0$ mocks (no contamination) and best-fitting theoretical prediction after accounting for the survey geometry and integral constraint effects. Bottom panel shows the residual power spectrum relative to the mean power spectrum. The dark and light shades represent the $68\%$ error on the mean and one realization, respectively. No imaging systematic cleaning is applied to these mocks.}\label{fig:model_mock}
\end{figure}

We employ a technique similar to the one proposed by \cite{chon2004fast} to account for the impact of the survey geometry on the theoretical power spectrum. The ensemble average for the partial sky power spectrum is related to that of the full sky power spectrum via a mode-mode coupling matrix, ${\rm M}_{\ell \ell^{\prime}}$,
\begin{equation}\label{eq:mixm}
    <\tilde{C}_{\ell}> = \sum_{\ell^{\prime}} {\rm M}_{\ell \ell^{\prime}}<C_{\ell^{\prime}}>.
\end{equation}
We convert this convolution in the spherical harmonic space into a multiplication in the correlation function space. Specifically, we first transform the theory power spectrum (Equation \ref{eq:cell}) to the correlation function, $\hat{\omega}^{\rm model}$. Then, we estimate the survey mask correlation function, $\hat{\omega}^{\rm window}$, and obtain the pseudo-power spectrum,
\begin{align}
    \tilde{C}^{\rm model}_{\ell} &= 2\pi \int \hat{\omega}^{\rm model}\hat{\omega}^{\rm window}~P_{\ell}(\cos\theta) d\cos\theta.
\end{align}
Figure \ref{fig:mask2pf} illustrates the survey mask correlation function ${\hat \omega}^{\rm window}$ for various masks representing the DESI footprint and its imaging sub-regions. Appendix \ref{ssec:windowconv} shows the validation of our method by comparing it to an alternative approach that computes the mode-mode coupling matrix ${\rm M}_{\ell \ell^{\prime}}$ and performs the convolution (Equation \ref{eq:mixm}) directly in $\ell$-space. We notice as the $\fnl$ deviates from zero, our approach introduces a noisy feature in the model, qualitatively in an unbiased manner ($\Delta \fnl < 1.1$). Figure \ref{fig:mcmc_mocks100} indeed demonstrates that our approach can recover the truth $\fnl$ in spite of the noisy feature.

The integral constraint is another systematic effect which is induced since the mean galaxy density is estimated from the observed galaxy density, and therefore is biased by the limited sky coverage \citep{peacock1991large}. To account for the integral constraint, the survey mask power spectrum is used to introduce a scale-dependent correction factor that needs to be subtracted from the power spectrum as,
\begin{equation}
     \tilde{C}^{\rm model, IC}_{\ell} = \tilde{C}^{\rm model}_{\ell} - \tilde{C}^{\rm model}_{\ell=0} \left(\frac{\tilde{C}^{\rm window}_{\ell}}{\tilde{C}^{\rm window}_{\ell=0}}\right),
\end{equation}
where $\tilde{C}^{\rm window}$ is the survey mask power spectrum, i.e., the spherical harmonic transform of $\hat{\omega}^{\rm window}$.

The lognormal simulations are used to validate the survey window and integral constraint correction. Figure \ref{fig:model_mock} shows the mean power spectrum of the $\fnl=0$ simulations (dashed) and the best-fitting theory prediction before and after accounting for the survey mask and integral constraint. The simulations are neither contaminated nor mitigated. The light and dark shades represent the 68\% estimated error on the mean and one single realization, respectively. The DESI mask, which covers around $40\%$ of the sky, is applied to the simulations. We find that the survey window effect modulates the clustering power on $\ell < 200$ and the integral constraint alters the clustering power on $\ell < 6$.

\subsection{Parameter estimation}

\begin{figure*}
\centering
\includegraphics[width=0.85\textwidth]{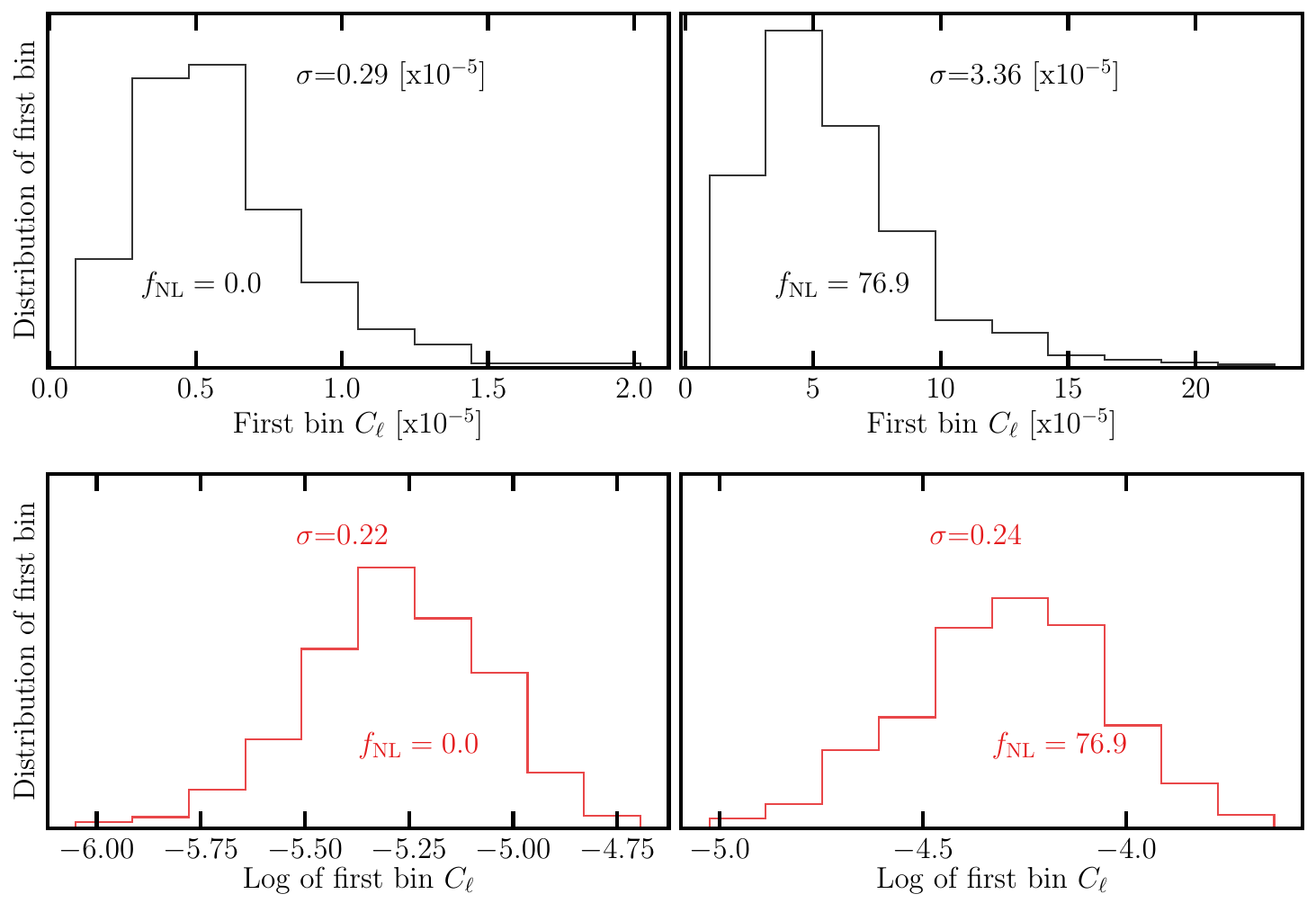}
\caption{The distribution of the first bin power spectra and its log transformation from the simulations with $\fnl=0$ (left) and $76.9$ (right). The log transformation largely removes the asymmetry in the distributions.}\label{fig:histcell}
\end{figure*}

Our parameter inference uses standard MCMC sampling. A constant clustering amplitude is assumed to determine the redshift evolution of the linear bias of the DESI LRG targets, $b(z) = b/D(z)$, which is supported by the HOD fits to the angular power spectrum \citep{zhou2021clustering}. In MCMC, we allow $\fnl$, $N_{\rm shot}$, and $b$ to vary, while all other cosmological parameters are fixed at the fiducial values (see \S \ref{ssec:mocks}). The galaxy power spectrum is divided into a discrete set of bandpower bins with $\Delta\ell=2$ between $\ell=2$ and $20$ and $\Delta \ell=10$ from $\ell=20$ to $300$. Each clustering mode is weighted by $2\ell+1$ when averaging over the modes in each bin.

The expected large-scale power is highly sensitive to the value of $\fnl$ such that the the amplitude of the covariance for $C_{\ell}$ is influenced by the true value of $\fnl$, see also \cite{2013MNRAS.428.1116R} for a discussion. As illustrated in the top row of Figure \ref{fig:histcell}, we find that the distribution of the power spectrum at the lowest bin, $2\leq \ell < 4$, is highly asymmetric and its standard deviation varies significantly from the simulations with $\fnl=0$ to $76.9$. We can make the covariance matrix less sensitive to $\fnl$ by taking the log transformation of the power spectrum, $\log C_{\ell}$. As shown in the bottom panels in Figure \ref{fig:histcell}, the log transformation reduces the asymmetry and the difference in the standard deviations between the $\fnl=0$ and $76.9$ simulations. Therefore, we minimize the negative log likelihood defined as,
\begin{equation}\label{eq:likelihood}
-2\log \mathcal{L} = (\log \tilde{C}(\Theta)-\log \tilde{C})^{\dagger} \mathbb{C}^{-1} (\log \tilde{C}(\Theta)-\log \tilde{C}),
\end{equation}
where $\Theta$ represents a container for the parameters $\fnl$, $b$, and $N_{\rm shot}$; $\tilde{C}(\Theta)$ is the (binned) expected pseudo-power spectrum; $\tilde{C}$ is the (binned) measured pseudo-power spectrum; and $\mathbb{C}$ is the covariance on $\log\tilde{C}$ constructed from the $\fnl=0$ log-normal simulations. Log-normal simulations have been commonly used and validated to estimate the covariance matrices for galaxy density fields, and non-linear effects are subdominant on the scales of interest to $\fnl$ \citep[see, e.g.,][]{2017MNRAS.466.1444C, 2021MNRAS.508.3125F}. We also test for the robustness of our results against an alternative covariance constructed from the $\fnl=76.9$ mocks. Flat priors are implemented for all parameters: $\fnl \in [-1000, 1000]$, $N_{\rm shot} \in [-0.001, 0.001]$, and $b \in [0, 5]$.

\subsection{Characterization of remaining systematics}\label{sec:systests}
\label{ssec:characterization}
One potential problem that can arise in the data-driven mitigation approach is \textit{over-correction}, which occurs when the corrections applied to the data remove the clustering signal and induce additional biases in the inferred parameter of interest. The neural network approach is more prone to this issue compared to the linear approach due to its increased degrees of freedom. As illustrated in the bottom panel of Figure \ref{fig:pcc}, the significant correlations among the imaging systematic maps may pose additional challenges for modeling the spurious fluctuations in the galaxy density field. Specifically, using highly correlated imaging systematic maps increases the statistical noise in the imaging weights, which elevates the potential for over subtracting the clustering power. These over-correction effects are estimated to have a negligible impact on baryon acoustic oscillations \citep{merz2021clustering}; however, they can significantly modulate the galaxy power spectrum on large scales, and thus lead to biased $\fnl$ constraints \citep{rezaie2021primordial, mueller2022primordial}. Although not explored thoroughly, the over-correction issues could limit the detectability of primordial features in the galaxy power spectrum and that of parity violations in higher order clustering statistics \citep{beutler2019primordial, cahn2021test, philcox2022probing}. Therefore, it is crucial to develop, implement, and apply techniques to minimize and control over-correction in the hope of ensuring that the constraints are as accurate and reliable as possible; one such approach is to reduce the dimensionality of the problem. Our goal is to reduce the correlations between the DESI LRG target density and the imaging systematic maps while controlling the over-correction effect. In the following, we describe how we approach this objective, by employing a series of simulations along with the residual systematics that we construct based on the cross power spectrum between the LRG density and imaging maps, and the mean LRG density as a function of imaging. We test different sets of the imaging systematic maps to identify the optimal set of the feature maps:
\begin{enumerate}[itemindent=*]
\item \textbf{Two maps}: Extinction, depth in z.
\item \textbf{Three maps}: Extinction, depth in z, psfsize in r.
\item \textbf{Four maps}: Extinction, depth in z, psfsize in r, stellar density.
\item \textbf{Eight maps}: Extinction, depth in $grzW1$, psfsize in $grz$.
\item \textbf{Nine maps}: Extinction, depth in $grzW1$, psfsize in $grz$, stellar density.
\item \textbf{Eleven maps}: same as Nine maps but with two additional maps; Extinction, depth in $grzW1$, psfsize in $grz$, stellar density, neutral hydrogen density, and photometric calibration in z.
\end{enumerate}
It is imperative to note that these sets are selected prior to examining the auto power spectrum of the LRG sample and unblinding the $\fnl$ constraints, and that the auto power spectrum and $\fnl$ measurements are unblinded only after our mitigation methods passed our rigorous tests for residual systematics. As detailed in the following, we discover that these tests tend to depend on the $\fnl$ value which is used in the mocks for the covariance matrix estimation.

\begin{figure*}
\centering
\includegraphics[width=0.95\textwidth]{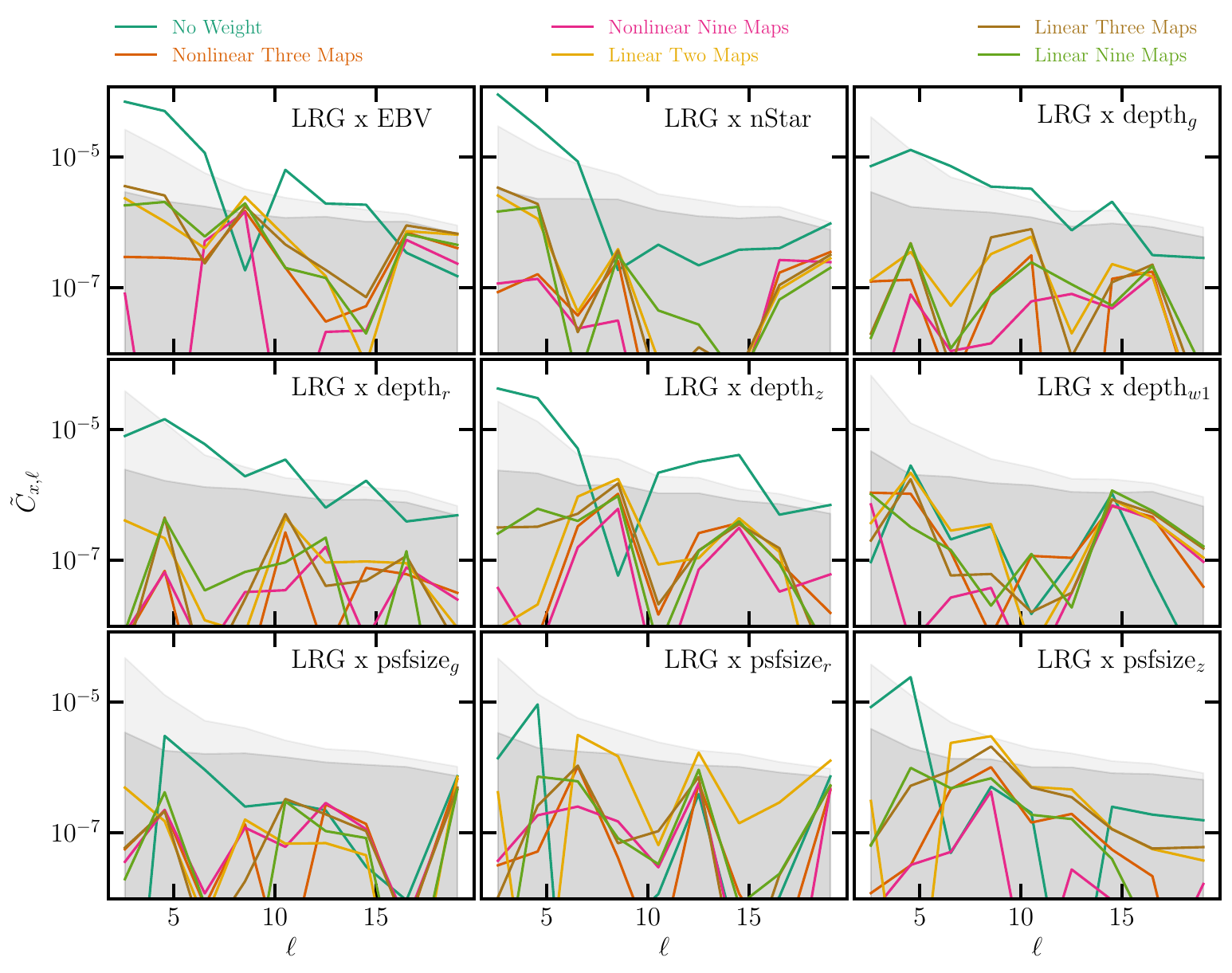}
\caption{The square of the cross power spectra between the DESI LRG targets and imaging systematic maps normalized by the auto power spectrum of the imaging systematic maps; see equation \ref{eq:cx}. The systematic maps considered are Galactic extinction (EBV), stellar density (nStar), depth in \textit{grzw1} (depth$_{grzw1}$), and seeing in \textit{grz} (psfsize$_{grz}$). The dark green curves display the cross spectra before imaging systematic correction (No Weight). The yellow, brown, and light green curves represent the results after applying the imaging weights from the linear models trained with \textit{two maps}, \textit{three maps}, and \textit{nine maps}. The orange and purple curves display the results after applying the imaging weights from the non-linear models trained with \textit{three maps} and \textit{nine maps}. The dark and light shades represent the $97.5$ percentile from cross correlating the imaging systematic maps and the $\fnl=0$ and $76.9$ lognormal density fields, respectively, without mitigation.}\label{fig:clxmock}
\end{figure*}

\begin{figure*}
\centering
\includegraphics[width=0.95\textwidth]{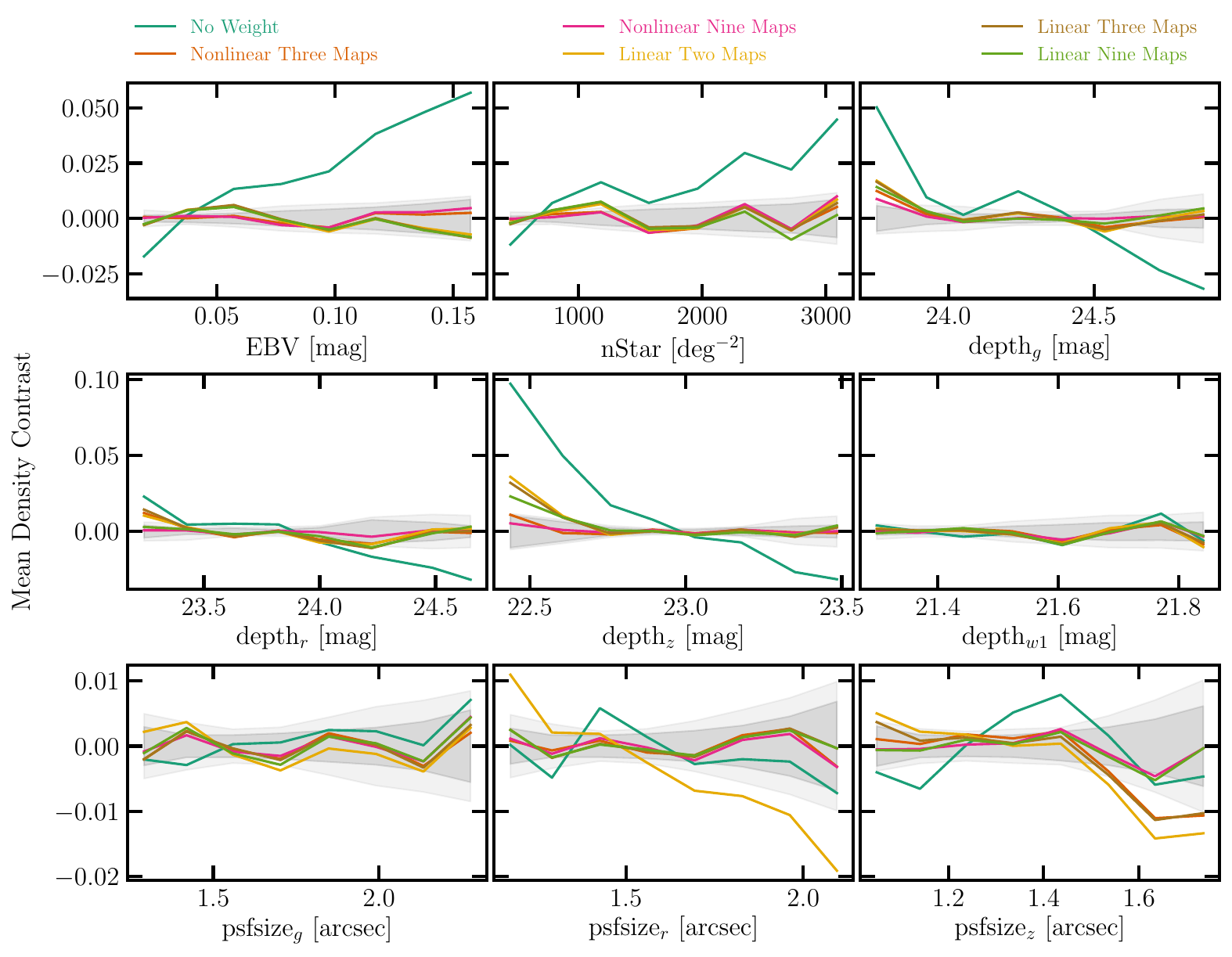}
\caption{The mean density contrast of the DESI LRG targets as a function of the imaging systematic maps: Galactic extinction (EBV), stellar density (nStar), depth in \textit{grzw1} (depth$_{grzw1}$), and seeing in \textit{grz} (psfsize$_{grz}$). The black curves display the results before imaging systematic correction. The red, blue and orange curves represent the relationships after applying the imaging weights from the linear models trained with \textit{two maps}, \textit{three maps}, and \textit{eight maps}, respectively. The green and pink curves display the results after applying the imaging weights from the non-linear models trained with \textit{three maps} and \textit{four maps}. The dark and light shades represent the $68\%$ dispersion of 1000 lognormal mocks with $\fnl=0$ and $76.9$, respectively.}\label{fig:nbarmock}
\end{figure*}

\subsubsection{Normalized cross power spectrum}

We characterize the cross correlations between the galaxy density and imaging systematic maps by
\begin{equation}
\tilde{C}_{X, \ell} = [\tilde{C}_{x_{1}, \ell}, \tilde{C}_{x_{2}, \ell}, \tilde{C}_{x_{3}, \ell}, ..., \tilde{C}_{x_{9}, \ell}],
\end{equation}
where $\tilde{C}_{x_{i}, \ell}$ represents the the square of the cross power spectrum between the galaxy density and $i^{\rm th}$ imaging map, $x_{i}$, divided by the auto power spectrum of $x_{i}$:
\begin{equation}\label{eq:cx}
\tilde{C}_{x_{i}, \ell} = \frac{(\tilde{C}_{gx_{i}, \ell})^{2}}{\tilde{C}_{x_{i}x_{i},\ell}}.
\end{equation}
With this normalization, $\tilde{C}_{x_{i}, \ell}$ estimates the contribution of systematics at every multipole up to the linear order to the galaxy power spectrum. Then, the $\chi^{2}$ value for the cross power spectra is calculated via,
\begin{equation}\label{eq:cx_chi2}
\chi^{2} = \tilde{C}^{T}_{X, \ell} \mathbb{C}_{X}^{-1} \tilde{C}_{X, \ell},
\end{equation}
where the covariance matrix $\mathbb{C}_{X} = < \tilde{C}_{X, \ell} \tilde{C}_{X, \ell'} >$ is constructed from the lognormal mocks. We consider both sets with $\fnl=0$ and $76.9$, and for each mitigated case, the covariance is from the mocks that have received the same treatment. These $\chi^{2}$ values are measured for every clean mock realization with the \textit{leave-one-out} technique and compared to the values observed in the DESI LRG targets with various imaging systematic corrections. Specifically, we use 999 realizations to estimate a covariance matrix and then apply the covariance matrix from the 999 realizations to measure the $\chi^{2}$ for the one remaining realization. This process is repeated for all 1000 realizations to construct a histogram for $\chi^{2}$. We only include the bandpower bins from $\ell=2$ to $20$ with $\Delta\ell=2$, which results in a total of 81 bins, and test for the robustness with higher $\ell$ modes in \ref{sec:scalesys}. 

Figure \ref{fig:clxmock} shows $\tilde{C}_{X,\ell}$ from the DESI LRG targets before and after applying various corrections for imaging systematics. The dark and light shades show the 97.5$^{\rm th}$ percentile from the $\fnl=0$ and $76.9$ mocks, respectively, that have had no mitigation applied to them. Without imaging weights (No Weight), the DESI LRG targets have the highest cross-correlations against extinction, stellar density, and depth in z. There are less significant correlations against depth in the g and r bands, and psfsize in the z band, which could be driven because of the inner correlations between the imaging systematic maps. First, we consider cleaning the DESI LRG targets with the linear model using two maps (extinction and depth in z) as identified from the Pearson correlation. Linear two maps is the least aggressive treatment method in terms of both the model flexibility and the number of input maps. With linear two maps, most of the cross correlation signals are reduced below statistical uncertainties, especially against extinction, stellar density, and depth. However, the cross correlations against psfsize in the r and z bands increase slightly on $6<\ell<20$ and $6<\ell<14$, respectively. This might be indicative of residual trends against psfsize.

The linear three maps approach alleviates the cross correlation against psfsize in r, and it yields similar results to those obtained from \textit{linear nine maps}, which indicates most of the contaminations can be attributed to these three maps. Therefore, we identify extinction, depth in z, and psfsize in r (\textit{three maps}) as the primary sources of systematic effects in the DESI LRG targets. Then, we adapt \textit{neural network three maps} to model non-linear systematic effects. Compared with the linear three maps method, we find that \textit{non-linear three maps} can reduce the cross correlations against both the r and z-band psfsize maps, which shows the benefit of using a non-linear approach. To further examine the robustness of our cleaning methods, we also show the cross correlations after cleaning the DESI LRG targets using nine imaging property maps (\textit{non-linear nine maps}). We do not find any significant residuals against the two extra maps for the neutral hydrogen density and photometric calibration in the z band. 

\subsubsection{Mean galaxy density contrast}

We calculate the histogram of the mean galaxy density contrast relative to the $j^{\rm th}$ imaging property, $x_{j}$:
\begin{equation}
\delta_{x_{j}} = ({\overline{\rho}})^{-1} \frac{\sum_{i} \rho_{i} W_{i}}{\sum_{i} W_{i}} - 1,
\end{equation}
where $\overline{\rho}$ is the global mean galaxy density, $W_{i}$ is the survey window in pixel $i$, and the summations over $i$ are evaluated from the pixels in every bin of $x_{j}$. We compute the histograms against all nine imaging properties (see Figure \ref{fig:ng}). We use a set of eight equal-width bins for every imaging map, which results in a total of 72 bins. Then, we construct the total mean density contract as,
\begin{equation}
\delta_{X} = [\delta_{x_{1}}, \delta_{x_{2}}, \delta_{x_{3}}, ..., \delta_{x_{9}}],
\end{equation}
and the total residual error as,
\begin{equation}
\chi^{2} = \delta_{X}^{T} \mathbb{C_{\delta}}^{-1} \delta_{X},
\end{equation}
where the covariance matrix $\mathbb{C}_{\delta} = < \delta_{X} \delta_{X}>$ is constructed from the lognormal mocks, in a consistent manner similar to the normalized cross power spectrum. Figure \ref{fig:nbarmock} shows the mean density contrast against the imaging properties for the DESI LRG targets. The dark and light shades represent the $1\sigma$ level fluctuations observed in 1000 lognormal density fields respectively with $\fnl=0$ and $76.9$ before mitigation. The DESI LRG targets before treatment (No Weight) exhibits a strong trend around $10\%$ against the z-band depth which is consistent with the cross power spectrum. Additionally, there are significant spurious trends against extinction and stellar density at about $5-6\%$. The linear approach is able to mitigate most of the systematic fluctuations with only extinction and depth in the z-band as input; however, a new trend appears against the r-band psfsize map with the \textit{linear two maps} approach, which is indicative of the psfsize-related systematics in the DESI LRG targets. This finding is in agreement with that from the cross power spectrum test. With linear three maps, we still observe around $2\%$ residual spurious fluctuations in the low end of depth in z and around $1\%$ in the high end of psfsize in z, which implies the presence of non-linear systematic effects. We find that the imaging weights from the non-linear model trained with the three identified maps (or four maps including the stellar density) are capable of reducing the fluctuations below $2\%$. Even with the non-linear three maps, we have about $1\%$ remaining systematic fluctuations against the z-band psfsize. The spurious trends are diminished especially when we adapt non-linear nine maps, especially against the low end of depth in g and r and against the high end of psfsize in z. 

\begin{figure*}
    \centering
    \includegraphics[width=\textwidth]{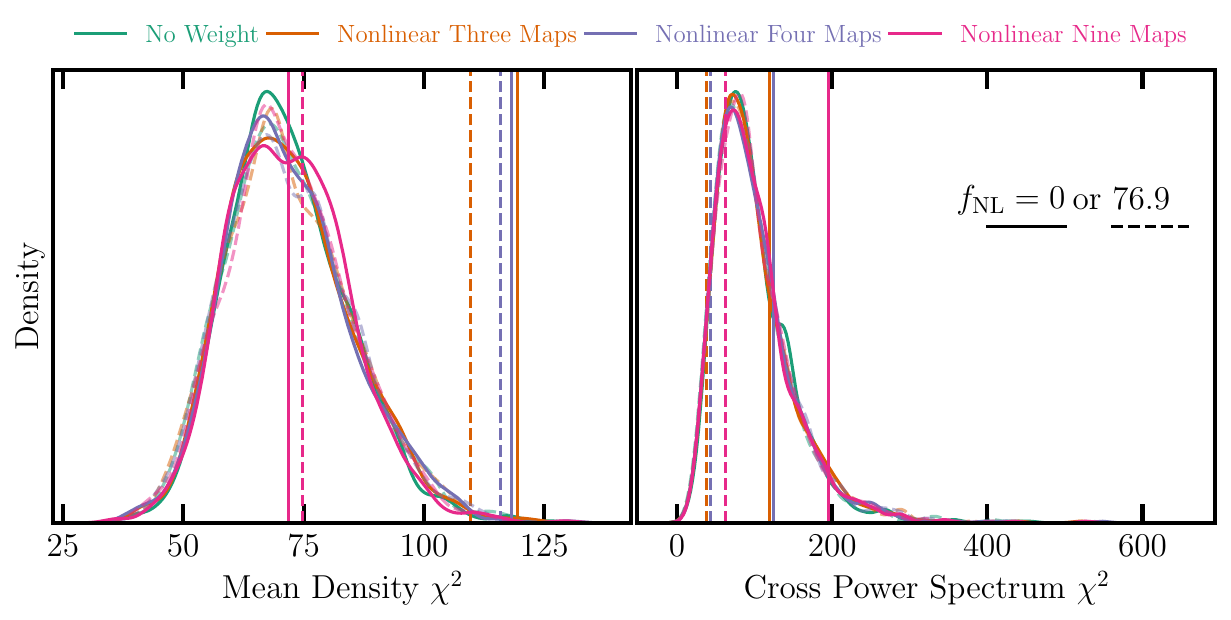}
    \caption{The remaining systematic error $\chi^{2}$ from the mean galaxy density contrast (left) and the galaxy-imaging normalized cross power spectrum (right). The values observed in the DESI LRG targets after the non-linear treatments are represented via vertical lines using the $\fnl=0$ (solid) or $76.9$ (dashed) covariance, and the histograms are constructed from 1000 realizations of clean lognormal mocks with $\fnl=0$ (solid) and $76.9$ (dashed).}\label{fig:chi2test}
\end{figure*}

\subsubsection{Residual error $\chi^{2}$}
We use the $\chi^{2}$ statistics to quantitatively assess how significant these mean density and cross power spectrum fluctuations are in comparison to the clean mocks. Figure \ref{fig:chi2test} presents $\chi^{2}$ histograms for the mean density contrast (left) and the normalized cross spectrum (right) statistics obtained from the lognormal mocks with different $\fnl$ values before and after applying mitigation methods. The mocks with $\fnl=0$ are shown with the solid curves while the other set with $\fnl=76.9$ are represented with the dashed curves. The use of the self-consistent covariance matrix (with respect to $\fnl$ or mitigation method) results in similar distributions, and therefore the mock histograms are employed as reference to evaluate the significance of residual systematics in the DESI LRG targets. We continue to use the self-consistent covariance, but consider both the $\fnl=0$ and $76.9$ covariance. The DESI LRG target $\chi^{2}$ values are compared via the vertical lines and summarized in Table \ref{tab:chi2test}. The solid and dashed vertical lines represent the values computed using the covariances based on the $\fnl=0$ and $76.9$ mocks, respectively. Regardless of the covariance used in the $\chi^{2}$ calculations, we find that the case without treatment (No Weight) exhibits serious contamination. For instance with the $\fnl=0$ covariance, the mean density and cross power errors are respectively $\chi^{2}=679.8$ and $20014.8$ (both with $p$-value $< 0.001$). These $\chi^{2}$ values are significantly high given that the degree of freedom is $72$ for the mean density and $81$ for the cross power spectrum. After cleaning, the $\chi^{2}$ values are decreased dramatically for both the mean density and normalized cross spectrum tests. The small impact on $\chi^{2}$ from including stellar density suggests that the stellar density trend can be explained by extinction due to the correlation between these properties, such that in regions with high stellar density, there is likely to be a higher concentration of dust, which can cause greater extinction of light. However, neither non-linear three maps nor non-linear four maps can reduce the mean density $\chi^{2}$ enough to be consistent with the mocks, indicating some significant residual error with $p$-values less than $0.005$.

The tests conducted here demonstrate the effectiveness of various cleaning approaches for the DESI LRG targets without revealing the measured power spectrum or $\fnl$ constraints. Overall, we observe that the non-linear method with the set of nine imaging property maps, successfully passes the mean density test irrespective of the covariance, as indicated by $\chi^{2}=71.9$ ($p$-value = $>0.48$) for the $\fnl=0$ covariance. On the other hand, the non-linear three maps and non-linear four maps methods both fail to sufficiently mitigate systematics in the mean density test, as evidenced by low $p$-values. Our work shows that it is essential to maintain a consistent covariance matrix, involving the same mitigation and ensuring consistency in $\fnl$ within the covariance. The sensitivity of the mean density $\chi^{2}$ to the $\fnl$ assumption in the covariance is notably lower, with greater reliance on the consistent mitigation method.  Conversely, the normalized cross spectrum $\chi^{2}$ exhibits a higher dependency on the $\fnl$ assumption in the covariance. The mean density diagnostic appears to be a less cosmology-sensitive probe of residual systematics. As a robustness test, we also increase the largest $\ell$ used in the $\chi^{2}$ calculation to $\ell=100$, which corresponds to density fluctuations on angles smaller than $2$ degrees. But we find no remaining systematic error from higher harmonic modes (see Appendix \ref{sec:scalesys}). The conclusion of these tests is that the non-linear method with the set of nine maps passes our null tests for the remaining systematics, and thus is chosen as the default approach for the treatment of imaging systematic effects. In the following subsection, we show how imaging systematic regressions remove clustering modes, with increasing dependence on the number of maps used, and thus bias the best fitting estimates of $\fnl$. Then, we present how we calibrate for the over-correction for our default mitigation method.

\begin{table*}
  \caption{Mean galaxy density contrast $\chi^{2}$ and normalized cross power spectrum $\chi^{2}$ from the DESI LRG targets and $p$-values that are inferred from the comparison to the $\fnl=0$ and $76.9$ clean mocks that have received the same mitigation. For the case of No Weight, we use the clean mocks without mitigation.}\label{tab:chi2test}
  \begin{tabular}{lcccc|cccc}
    \hline
    \hline
    \multirow{3}{*}{} &&
      \multicolumn{2}{c}{\textbf{Mean Density Contrast (dof=72)}} & & &
      \multicolumn{2}{c}{\textbf{Cross Power Spectrum (dof=81)}} \\
    & \multicolumn{2}{c}{Covariance: $\fnl$=0} & \multicolumn{2}{c}{$\fnl$=76.9} & \multicolumn{2}{c}{$\fnl$=0} & \multicolumn{2}{c}{$\fnl$=76.9} \\
    \hline
    \textbf{Method} & $\chi^{2}$ & $p$-value & $\chi^{2}$ & $p$-value & $\chi^{2}$ & $p$-value & $\chi^{2}$ & $p$-value \\
    \hline
No Weight & 679.8 & < 0.001 & 405.2 & < 0.001 & 20014.8 & < 0.001 & 721.1 & < 0.001 \\
Nonlinear Three Maps & 119.5 & 0.002 & 109.7 & 0.003 & 118.6 & 0.273 & 38.0 & 0.951 \\
Nonlinear Four Maps & 118.2 & 0.001 & 115.9 & 0.001 & 124.6 & 0.240 & 43.3 & 0.921 \\
Nonlinear Nine Maps & 71.9 & 0.487 & 74.9 & 0.392 & 195.1 & 0.047 & 62.2 & 0.767 \\
    \hline
  \end{tabular}
\end{table*}

\subsection{Calibration of over-correction}\label{ssec:calibration}

The template-based mitigation of imaging systematics removes some of the true clustering signal, and mitigating with more maps should remove more modes and thus bias both the $\fnl$ estimation and its associated uncertainty. We calibrate the over-correction effect using the mocks presented in \S \ref{sec:data}. Having two sets of mocks with low and high power at large scales (low $\ell$) offers a key advantage: it provides a model for mapping the entire posterior distribution, which enables sus to understand how the constraints on $\fnl$ degrade as the magnitude of the imaging systematic correction increases.  We apply the neural network model to both the $\fnl=0$ and $76.9$ simulations, with and without imaging systematics, using various sets of imaging systematic maps. Specifically, we consider \textit{non-linear three maps}, \textit{non-linear four maps}, and \textit{non-linear nine maps}. Then, we measure the power spectra from the mocks. We fit both the mean power spectrum and each individual power spectrum from the mocks. Appendix \ref{ssec:contmocks} outlines the impact of the non-linear methods on the mock power spectra, and here we summarize relevant details for the calibration of over-correction.

\begin{figure}
\centering
\includegraphics[width=0.45\textwidth]{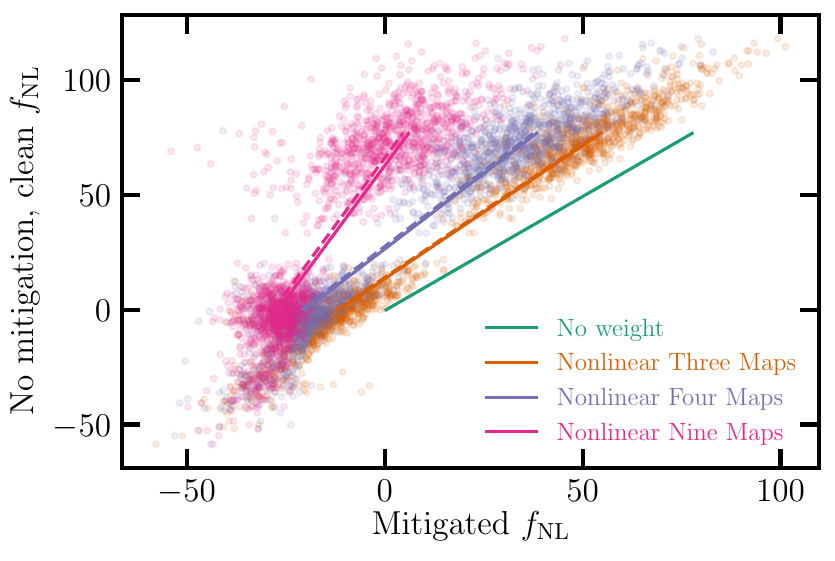}
\caption{The \textit{No mitigated, clean} vs \textit{mitigated} $\fnl$ values from the $\fnl=0$ and $76.9$ mocks. The solid (dashed) lines represent the best-fitting estimates from fitting the mean power spectrum of the clean (contaminated) mocks. The scatter points show the best-fitting estimates from fitting the individual spectra for the clean mocks.}\label{fig:fnlbias}
\end{figure}

Fihgure \ref{fig:fnlbias} displays a comparison between the best-fitting estimates of $\fnl$ before and after mitigation for the clean mocks. The best-fitting estimates from the mean of the mocks are represented by the solid curves, and the individual spectra results are displayed as the scatter points. The results from fitting the mean power spectrum of the contaminated mocks are also shown via the dashed curves. We find nearly identical results for the biases caused by mitigation, whether or not the mocks have any contamination, which can be seen by observing the solid and dashed curves displayed on Figure \ref{fig:fnlbias} (see, also, Figure \ref{fig:clmocks}, for a comparison of the mean power spectrum). For clarity, the best-fitting estimates for the individual contaminated data are not shown in the figure.

As summarized in Table \ref{tab:contmocksmcmc}, we observe notable shifts in the best-fitting estimates of $\fnl$ obtained from the mean power spectrum of the mocks. Specifically, for the $\fnl=0$ mocks, we obtain $\Delta \fnl=-12$ for non-linear three maps, $-20$ for non-linear four maps, and $-27$ for non-linear nine maps. Larger shifts are evident for $\fnl=76.9$: $\Delta \fnl=-23$ for non-linear three maps, $-39$ for non-linear four maps, and $-72$ for non-linear nine maps. These factor imply that the effect of systematic mitigation on the inferred $\fnl$ depends on the true value of $\fnl$.

To calibrate our methods, we fit a linear curve to the $\fnl$ estimates from the mean power spectrum of the mocks, $f_{\rm NL, no~mitigation, clean} = m_{1} f_{\rm NL, mitigated} + m_{2}$. The $m_{1}$ and $m_{2}$ coefficients for non-linear three, four, and nine maps are summarized in Table \ref{tab:debiasparams}. These coefficients represent the impact of the cleaning methods on the likelihood. The uncertainty in $\fnl$ after mitigation increases by $m_{1}-1$. Figure \ref{fig:fnlbias} also shows that the choice of our cleaning method can have significant implications for the accuracy of the measured $\fnl$, and careful consideration should be given to the selection of the primary imaging systematic maps and the calibration of the cleaning algorithms in order to minimize systematic uncertainties.

\begin{table}
\begin{center}
\caption{Linear parameters employed to de-bias the $\fnl$ constraints to account for the over-correction issue.}\label{tab:debiasparams}
\begin{tabular}{lcc}
\hline
\hline
\textbf{Cleaning Method} & $m_{1}$ & $m_{2}$ \\
\hline
Nonlinear Three Maps & 1.17 & 13.95 \\
Nonlinear Four Maps & 1.32 & 26.97 \\
Nonlinear Nine Maps & 2.35 & 63.5\\
\hline
\end{tabular}
\end{center}
\end{table}

%% file: sections/results.tex
\section{Results}\label{sec:results}
We now present our $\fnl$ constraints obtained from the power spectrum of the DESI LRG targets. The treatment of the imaging systematic effects is performed on each imaging region (BASS+MzLS, DECaLS North/South) separately. After cleaning, the regions are combined for the measurement of the power spectrum. We unblind the galaxy power spectrum and the $\fnl$ values after our cleaning methods are validated and vetted by the cross power spectrum and mean galaxy density diagnostics. As presented in Section \ref{sec:systests}, these tests show that none of the linear methods yields reasonable statistics, and only the nonlinear approach with the nine maps can pass the criteria, which is why we select the nonlinear nine maps as our fiducial method for cleaning systematics. We also conduct additional tests to check the robustness of our constraints against various assumptions, such as analyzing each region separately, applying cuts on imaging conditions, and changing the smallest mode used in fitting for $\fnl$.

\subsection{DESI imaging LRG sample}
\begin{figure}
    \centering
    \includegraphics[width=0.47\textwidth]{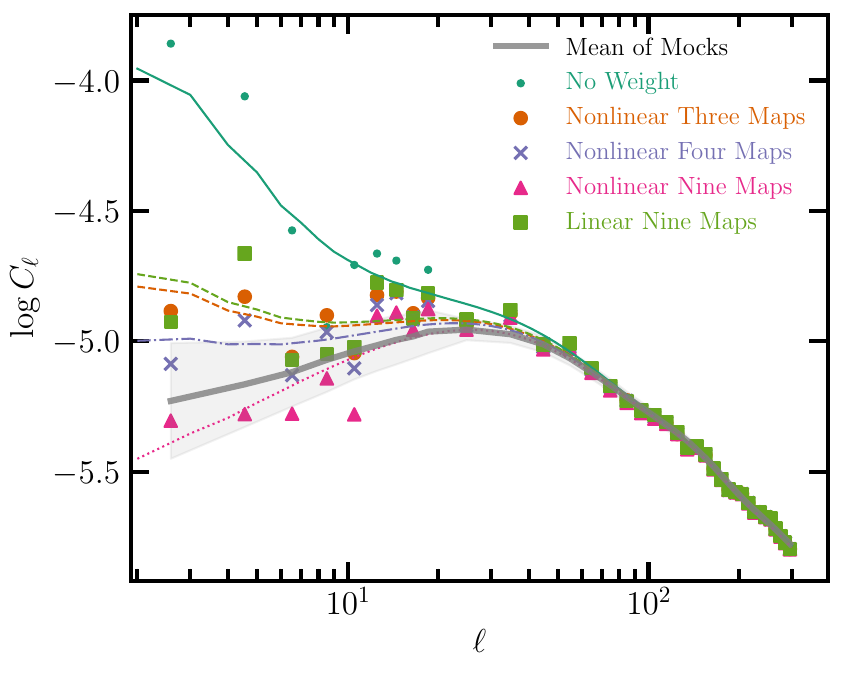} 
    \caption{The angular power spectrum of the DESI LRG targets before (\textit{No weight}) and after correcting for imaging systematics using the linear and non-linear methods. The curves represent the corresponding best-fitting theory predictions. The solid curve and grey shade respectively represent the mean power spectrum and $68\%$ error from the $\fnl=0$ mocks.}
    \label{fig:cl_dr9}
\end{figure}

We find that the excess clustering signal in the power spectrum of the DESI LRG targets is mitigated after correcting for the imaging systematic effects. Figure \ref{fig:cl_dr9} shows the measured power spectrum of the DESI LRG targets before and after applying imaging weights and the best-fitting theory curves. The solid grey line and the grey shade represent respectively the mean power spectrum and 1$\sigma$ error, estimated from the $\fnl=0$ lognormal simulations. The differences between various cleaning methods are significant on large scales ($\ell < 20$), but the small scale clustering measurements are consistent. We associate the differences to the over-correction caused by including more maps for the treatment of systematics, which we base upon the validation of the methods on the mocks, or the suppression of excess power from systematics. Comparing \textit{non-linear three maps} to \textit{non-linear four maps}, we find that adding stellar density in the non-linear approach (\textit{non-linear four maps}) further reduces the excess power relative to the mock power spectrum, in particular on modes between $2\leq \ell < 4$. However, when calibrated on the lognormal simulations, we find that the over-subtraction due to stellar density is reversed after accounting for over-correction. Our fiducial approach, \textit{non-linear nine maps}, yields the lowest (and almost constant) power on large scales among all methods.

\subsubsection{Calibrated constraints}

\begin{figure}
    \raggedleft
    \includegraphics[width=0.439\textwidth, trim={0 1.4cm 0 0},clip]{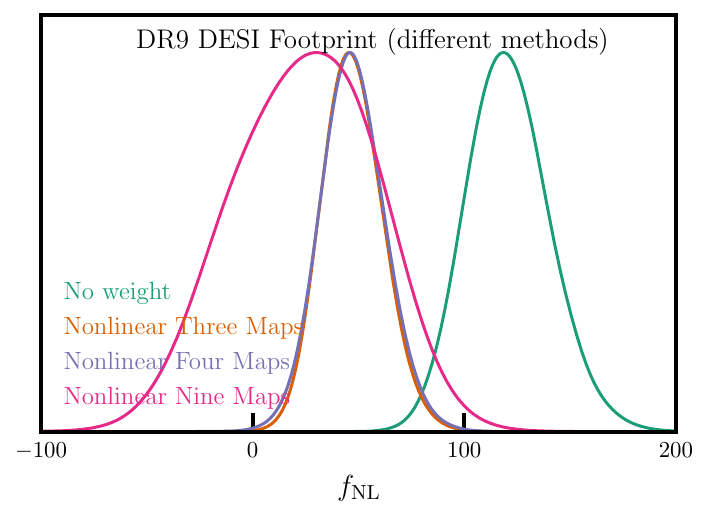}
    \includegraphics[width=0.47\textwidth, trim={0 0 0.15cm 0.2cm},clip]{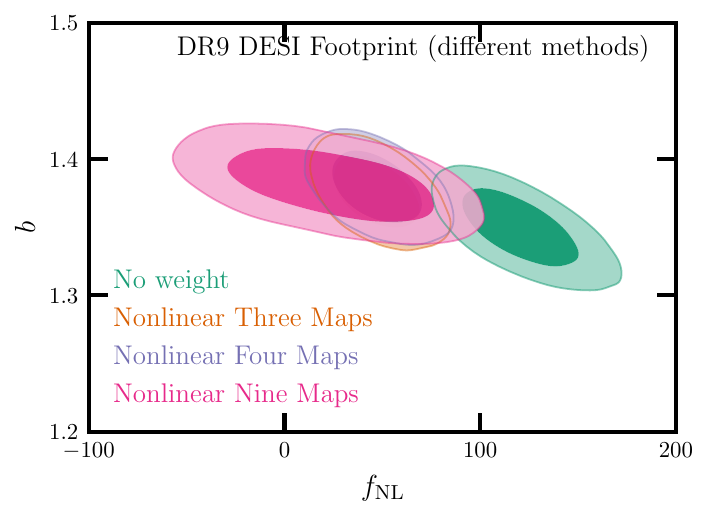} 
    \caption{The calibrated constrains from the DESI LRG targets. \textit{Top}: probability distribution for $\fnl$ marginalized over the shotnoise and bias. \textit{Bottom}: $68\%$ and $95\%$ probability distribution contours for the bias and $\fnl$ from the DESI LRG targets before and after applying the non-linear cleaning methods. The lognormal mocks are used to calibrate these distributions for over-correction.}\label{fig:mcmc_dr9}
\end{figure}

\begin{figure}
\raggedleft
\includegraphics[width=0.45\textwidth]{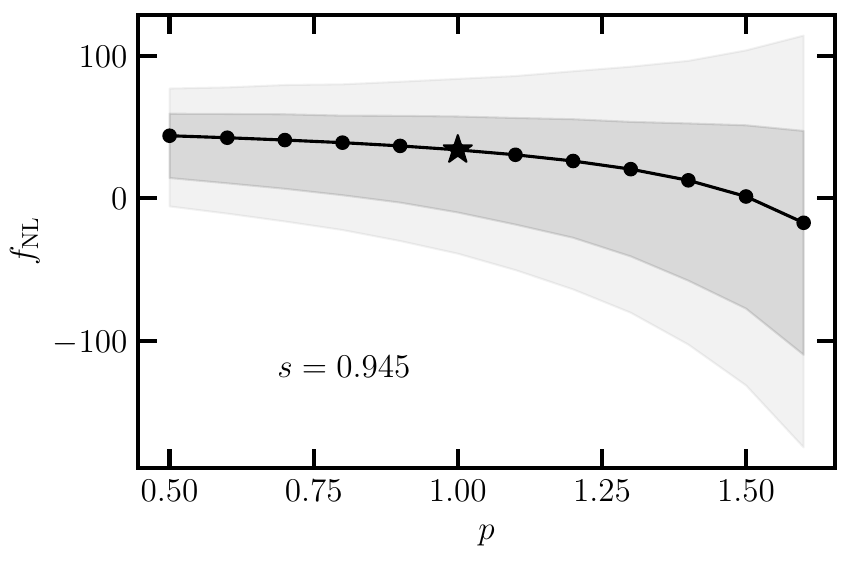}
\includegraphics[width=0.437\textwidth]{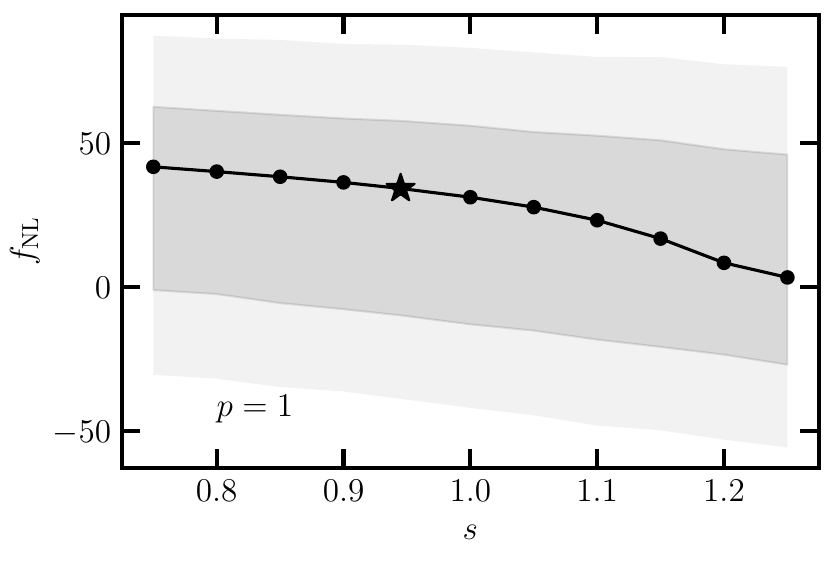}
\caption{The best-fitting estimates of $\fnl$ and their corresponding $68\%$ ($95\%$) errors from the DESI LRG targets using the non-linear nine maps approach given various values of $p$ or $s$. The star symbol represents the fiducial analysis with $p=1$ and $s=0.945$.}\label{fig:fnl_magbias}
\end{figure}

\begin{table*}
    \caption{The calibrated best-fitting, marginalized mean, and marginalized $68\%$ ($95\%$) confidence estimates for $\fnl$ from fitting the power spectrum of the DESI LRG targets before and after correcting for imaging systematic effects. The lowest mode is $\ell_{\rm min}=2$, $p=1$, and $s=0.945$.}
    \label{tab:dr9methodcalib}
   \centerline{%
    \begin{tabular}{llllllll}
    \hline
    \hline
   &  & 	  & & $\fnl$ &  &  \\
   \cmidrule(r{.7cm}){3-6}
Footprint      & Method & 	Best fit  & Mean & $ 68\%$ CL & $ 95\%$ CL & $\chi^{2}$ (dof=$34$) \\
    \hline
DESI & No Weight                        & $   118$& $   121$& $   102<\fnl<   140$& $    86<\fnl<   161$ &   45.1\\
DESI & Nonlinear Three Maps             & $    46$& $    47$& $    33<\fnl<    61$& $    21<\fnl<    76$ &   33.9\\
DESI & Nonlinear Four Maps              & $    46$& $    47$& $    33<\fnl<    62$& $    19<\fnl<    78$ &   34.4\\
DESI & Nonlinear Nine Maps              & $    34$& $    24$& $   -10<\fnl<    58$& $   -39<\fnl<    84$ &   39.1\\    
   \hline
    \end{tabular}
}
\end{table*}

All $\fnl$ constraints presented here are calibrated for the effect of over-correction using the lognormal simulations. Table \ref{tab:dr9methodcalib} describes the best-fitting and marginalized mean estimates of $\fnl$ from fitting the power spectrum of the DESI LRG targets before and after cleaning with the non-linear approach given various combinations for the imaging systematic maps. Figure \ref{fig:mcmc_dr9} shows the marginalized probability distribution for $\fnl$ in the top panel, and the $68\%$ and $95\%$ probability contours for the linear bias parameter and $\fnl$ in the bottom panel, from our sample before and after applying various corrections for imaging systematics. Overall, we find the maximum likelihood estimates to be consistent among the various cleaning methods. We obtain $33 (21) < \fnl < 61(76)$ at $68\%(95\%)$ confidence with $\chi^{2}=33.9$ for \textit{non-linear three maps} with $34$ degrees of freedom. Accounted for over-correction, we obtain $33(19) < \fnl < 62(78)$ with $\chi^{2}=34.4$ using \textit{non-linear four maps} which includes the additional stellar density map. With or without stellar density, the confidence intervals are consistent with each other and significantly off from zero PNG; specifically, the probability that $\fnl$ is erroneously greater than zero, $P(\fnl >0)=99.9$ per cent, which we attribute to systematics (see Section \ref{sec:systests}). We also apply a more aggressive systematics treatment that includes regression using the non-linear approach against the set of nine imaging maps we identified, \textit{non-linear nine maps}, and find that zero $\fnl$ is recovered. Specifically, our maximum likelihood value changes to $\fnl \sim 34$ with $\chi^{2}=39.1$, and the uncertainty on $\fnl$ increases by more than a factor of two, resulting in $-10 (-39) < \fnl < 58 (84)$ at $68\% (95\%)$ confidence. This increase is attributed to the aggressive treatment, which removes large-scale clustering information and diminishes the constraining power of the dataset.

Additionally, we explore the sensitivity of the $\fnl$ posterior using the non-linear nine maps method while varying the values of $p$ in the range of $0.5$ to $1.6$ and $s$ in the range of $0.75$ to $1.25$. Figure \ref{fig:fnl_magbias} illustrates our findings, and Table \ref{tab:dr9ps} provides a summary. Regardless of the specific values chosen for $p$ and $s$, we reliably recover $\fnl=0$ within the $95\%$ confidence interval. The top panel also implies that marginalizing over $p$ can induce projection effects and lead to biased $\fnl$ constraints. For comparison, we obtain $102(86) < \fnl < 140(161)$ at $68\% (95\%)$ confidence with $\chi^{2}=45.1$ for the \textit{no weight} case.

\begin{table*}
    \caption{The calibrated best-fitting, marginalized mean, and marginalized $68\%$ ($95\%$) confidence estimates for $\fnl$ from the DESI LRG targets cleaned with the non-linear nine maps approach, given various values of $p$ and $s$. The fiducial analysis uses $p=1$ and $s=0.945$ (DESI footprint).}
    \label{tab:dr9ps}
   \centerline{%
    \begin{tabular}{llllll}
    \hline
    \hline
   &  & 	  & $\fnl$ &  &    \\
   \cmidrule(r{.7cm}){2-5}
Parameter & 	Best fit  & Mean & $ 68\%$ CL & $ 95\%$ CL & $\chi^{2}$ (dof=$34$) \\
\hline
$p=$ 0.5                                     & $    44$& $    37$& $    14<\fnl<    60$& $    -6<\fnl<    77$ &   39.1\\
0.6                                     & $    43$& $    35$& $    11<\fnl<    59$& $   -11<\fnl<    78$ &   39.1\\
0.7                                     & $    41$& $    33$& $     7<\fnl<    59$& $   -16<\fnl<    80$ &   39.1\\
0.8                                     & $    39$& $    31$& $     2<\fnl<    58$& $   -22<\fnl<    80$ &   39.1\\
0.9                                     & $    37$& $    28$& $    -3<\fnl<    58$& $   -30<\fnl<    82$ &   39.1\\
1.0                                     & $    34$& $    24$& $   -10<\fnl<    58$& $   -39<\fnl<    84$ &   39.1\\
1.1                                     & $    31$& $    19$& $   -18<\fnl<    57$& $   -50<\fnl<    86$ &   39.1\\
1.2                                     & $    26$& $    15$& $   -28<\fnl<    56$& $   -64<\fnl<    89$ &   39.1\\
1.3                                     & $    21$& $     7$& $   -41<\fnl<    54$& $   -80<\fnl<    93$ &   39.0\\
1.4                                     & $    13$& $    -2$& $   -58<\fnl<    53$& $  -103<\fnl<    97$ &   39.0\\
1.5                                     & $     1$& $   -13$& $   -77<\fnl<    51$& $  -131<\fnl<   104$ &   39.0\\
1.6                                     & $   -17$& $   -31$& $  -110<\fnl<    47$& $  -175<\fnl<   114$ &   39.0\\
\hline
$s=$ 0.75                                    & $    42$& $    31$& $    -1<\fnl<    62$& $   -30<\fnl<    87$ &   39.2\\
0.80                                    & $    40$& $    30$& $    -3<\fnl<    61$& $   -32<\fnl<    86$ &   39.1\\
0.85                                    & $    38$& $    28$& $    -6<\fnl<    60$& $   -35<\fnl<    86$ &   39.1\\
0.90                                    & $    36$& $    26$& $    -8<\fnl<    58$& $   -36<\fnl<    84$ &   39.1\\
0.945                                   & $    34$& $    24$& $   -10<\fnl<    58$& $   -39<\fnl<    84$ &   39.1\\
1.00                                    & $    31$& $    22$& $   -13<\fnl<    56$& $   -42<\fnl<    83$ &   39.0\\
1.05                                    & $    28$& $    19$& $   -15<\fnl<    54$& $   -45<\fnl<    81$ &   39.0\\
1.10                                    & $    23$& $    17$& $   -18<\fnl<    52$& $   -48<\fnl<    80$ &   39.0\\
1.15                                    & $    17$& $    15$& $   -21<\fnl<    51$& $   -50<\fnl<    80$ &   38.9\\
1.20                                    & $     8$& $    12$& $   -24<\fnl<    48$& $   -53<\fnl<    77$ &   38.9\\
1.25                                    & $     3$& $     9$& $   -27<\fnl<    46$& $   -56<\fnl<    76$ &   38.8\\
    \end{tabular}}
\end{table*}

\subsubsection{Uncalibrated constraints: robustness tests}
\begin{figure}
    \centering
    \includegraphics[width=0.45\textwidth]{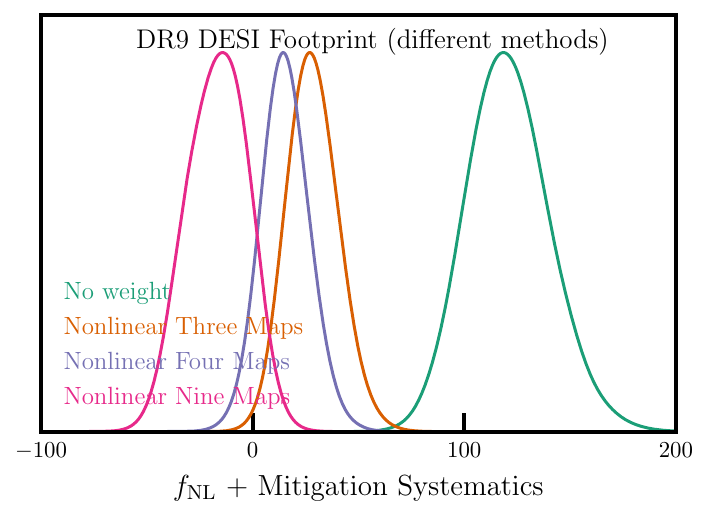}
    \caption{Same as Figure \ref{fig:mcmc_dr9} but without accouting for over-correction. }
    \label{fig:mcmcdr9noshift}
\end{figure}

Figure \ref{fig:mcmcdr9noshift} shows the probability distributions of $\fnl$ for various treatments before accounting for the over-correction effect. The method with the largest flexibility and more number of imaging systematic maps is more likely to regress out the clustering signal aggressively and return biased $\fnl$ constraints. The non-linear three maps approach returns a best-fitting estimate of $\fnl=27$ with the $68\%(95\%)$ confidence of $17(6)<\fnl <40(53)$ and $\chi^{2}=33.9$. With the stellar density map included, non-linear four maps yields a smaller best-fitting estimates of $\fnl=14$ with the error of $5(-6)<\fnl<26(38)$. The non-linear nine maps gives an asymmetric posterior with the marginalized mean $\fnl=-17$, and the smallest best-fitting estimate of $\fnl=-13$ with the error of $-31(-44)<\fnl<-3(9)$. The disparities in the best-fitting estimates can be linked to over-correction, mirroring the effects observed in the mocks (refer to Figure \ref{fig:contmcmc}). Consequently, caution is advised when considering the uncalibrated values. Without adjusting for over-correction, non-linear four maps and non-linear nine maps recover zero $\fnl$ within $95\%$ and $68\%$ confidence, respectively. However, the non-linear method with three maps exhibits tension with $\fnl=0$ at a confidence level of $99.5$ percent.

\begin{table*}
    \caption{The uncalibrated best-fitting and marginalized mean estimates for $\fnl$ from fitting the power spectrum of the DESI LRG targets before and after correcting for systematics. The estimates are not calibrated for over-correction, and thus are subject to mitigation systematics. The number of degrees of freedom is 34 (37 data points - 3 parameters) for all cases except the case that combines the data at the likelihood level, `BASS+MzLS+DECaLS', in which the dof is 104 ($3\times37-7$). The lowest mode is $\ell=2$ and the covariance matrix is from the $\fnl=0$ clean mocks (no mitigation) except for the case with '+ Cov' in which the covariance matrix is from the $\fnl=76.9$ clean mocks (no mitigation). We fix $p=1$ for all cases and $s=0.945$ for DESI, $0.943$ for DECaLS North (and South), and $0.951$ for BASS+MzLS.}
    \label{tab:dr9method}
    \begin{tabular}{lllllll}
    \hline
    \hline
   &  & &  &  $\fnl$ + Mitigation Systematics   & &  \\
   \cmidrule(r{.7cm}){3-6}
Footprint                               & Method & 	Best fit  & Mean & $ 68\%$ CL & $ 95\%$ CL & $\chi^{2}$ (dof=$34$) \\
    \hline
\bf{DESI} & \bf{No Weight}                        & $   \bf{118}$& $   \bf{121}$& $   \bf{102}<\fnl<   \bf{140}$& $    \bf{86}< \fnl<   \bf{161}$ &   \bf{45.1}\\
DESI & Linear Three Maps                & $    36$& $    37$& $    25<\fnl<    50$& $    14<\fnl<    64$ &   38.6\\
DESI & Linear Four Maps                 & $    31$& $    32$& $    20<\fnl<    45$& $     9<\fnl<    58$ &   40.3\\
\bf{DESI} & \bf{Linear Nine Maps}                 & $    \bf{30}$& $    \bf{32}$& $    \bf{19}<\fnl<   \bf{43}$& $     \bf{9}<\fnl<  \bf{57}$ &   \bf{41.9}\\
DESI & Nonlinear Three Maps             & $    27$& $    28$& $    17<\fnl<    40$& $     6<\fnl<    53$ &   33.9\\
DESI & Nonlinear Four Maps              & $    14$& $    15$& $     5<\fnl<    26$& $    -6<\fnl<    38$ &   34.4\\
\bf{DESI} & \bf{Nonlinear Nine Maps}              & $   \bf{-13}$& $   \bf{-17}$& $   \bf{-31}<\fnl<    \bf{-3}$& $   \bf{-44}<\fnl<     \bf{9}$ &  \bf{39.1}\\
DESI (imag. cut) & Nonlinear Nine Maps  & $   -25$& $   -22$& $   -37<\fnl<    -7$& $   -49<\fnl<     6$ &   37.7\\
DESI (comp. cut) & Nonlinear Nine Maps  & $   -24$& $   -23$& $   -35<\fnl<   -10$& $   -46<\fnl<     2$ &   36.3\\
DESI & Nonlinear Nine Maps+$f_{\rm NL}=76.9$ Cov& $   -11$& $   -15$& $   -30<\fnl<    0$& $   -43<\fnl<    12$ &   37.4\\
BASS+MzLS+DECaLS & Nonlinear Nine Maps& $   -31$& $   -26$& $   -41<\fnl<    -9$& $   -53<\fnl<     5$ &  114.2\\
   \hline
BASS+MzLS & Nonlinear Three Maps        & $    13$& $    16$& $    -6<\fnl<    38$& $   -28<\fnl<    64$ &   34.9\\
BASS+MzLS & Nonlinear Four Maps         & $    10$& $    12$& $   -11<\fnl<    34$& $   -35<\fnl<    59$ &   34.1\\
\bf{BASS+MzLS} & \bf{Nonlinear Nine Maps}         & $    \bf{-9}$& $   \bf{-13}$& $   \bf{-37}<\fnl<    \bf{10}$& $   \bf{-59}<\fnl<    \bf{32}$ &   \bf{36.4}\\
BASS+MzLS (imag. cut) & Nonlinear Nine Maps& $   -12$& $   -13$& $   -36<\fnl<    10$& $   -58<\fnl<    34$ &   36.7\\
BASS+MzLS (comp. cut) & Nonlinear Nine Maps& $   -15$& $   -16$& $   -39<\fnl<     6$& $   -61<\fnl<    28$ &   35.3\\
   \hline
DECaLS North & Nonlinear Three Maps     & $    41$& $    45$& $    21<\fnl<    69$& $    -1<\fnl<    98$ &   40.8\\
DECaLS North & Nonlinear Four Maps      & $    30$& $    32$& $    10<\fnl<    56$& $   -18<\fnl<    83$ &   40.9\\
\bf{DECaLS North} & \bf{Nonlinear Nine Maps}      & $    \bf{-4}$& $ \bf{-13}$& $  \bf{-40}<\fnl<   \bf{13}$& $   \bf{-64}<\fnl<   \bf{36}$ &  \bf{44.6}\\
DECaLS North (imag. cut) & Nonlinear Nine Maps& $   -16$& $   -20$& $   -47<\fnl<     7$& $   -70<\fnl<    31$ &   36.1\\
DECaLS North (comp. cut) & Nonlinear Nine Maps& $   -17$& $   -20$& $   -46<\fnl<     5$& $   -68<\fnl<    28$ &   42.7\\
DECaLS North (no DEC cut) & Nonlinear Nine Maps& $     0$& $   -13$& $   -43<\fnl<    15$& $   -67<\fnl<    38$ &   44.2\\
DECaLS North & Nonlinear Eleven Maps        & $    -2$& $    -7$& $   -32<\fnl<    16$& $   -54<\fnl<    39$ &   40.0\\
   \hline
DECaLS South & Nonlinear Three Maps     & $    30$& $    31$& $    11<\fnl<    53$& $   -28<\fnl<    76$ &   30.2\\
DECaLS South & Nonlinear Four Maps      & $   -42$& $    -5$& $   -44<\fnl<    27$& $   -70<\fnl<    49$ &   33.4\\
\bf{DECaLS South} & \bf{Nonlinear Nine Maps}      & $   \bf{-43}$& $  \bf{-40}$& $   \bf{-58}<\fnl<  \bf{-21}$& $   \bf{-75}<\fnl<    \bf{ 3}$ &  \bf{31.3}\\
DECaLS South (imag. cut) & Nonlinear Nine Maps& $   -57$& $   -55$& $   -76<\fnl<   -36$& $   -96<\fnl<    -8$ &   30.0\\
DECaLS South (comp. cut) & Nonlinear Nine Maps& $   -42$& $   -40$& $   -58<\fnl<   -22$& $   -76<\fnl<    -1$ &   30.4\\
DECaLS South (no DEC cut) & Nonlinear Nine Maps& $    -2$& $   -10$& $   -31<\fnl<    10$& $   -50<\fnl<    26$ &   26.1\\
DECaLS South & Nonlinear Eleven Maps        & $   -38$& $   -35$& $   -52<\fnl<   -16$& $   -70<\fnl<     5$ &   32.3\\
   \hline
    \end{tabular}
\end{table*}

\begin{figure}
    \centering
    \includegraphics[width=0.45\textwidth]{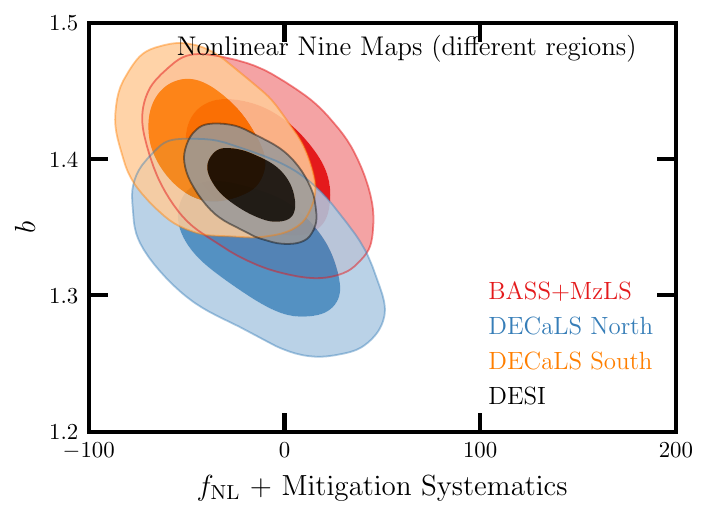} 
    \caption{The uncalibrated 2D constraints from the DESI LRG targets using the nonlinear nine maps treatment for each imaging survey compared with that for the whole DESI footprint. The dark and light shades represent the $68\%$ and $95\%$ confidence intervals, respectively.}\label{fig:mcmc_dr9reg}
\end{figure}
\begin{figure*}
    \centering
    \includegraphics[width=0.9\textwidth]{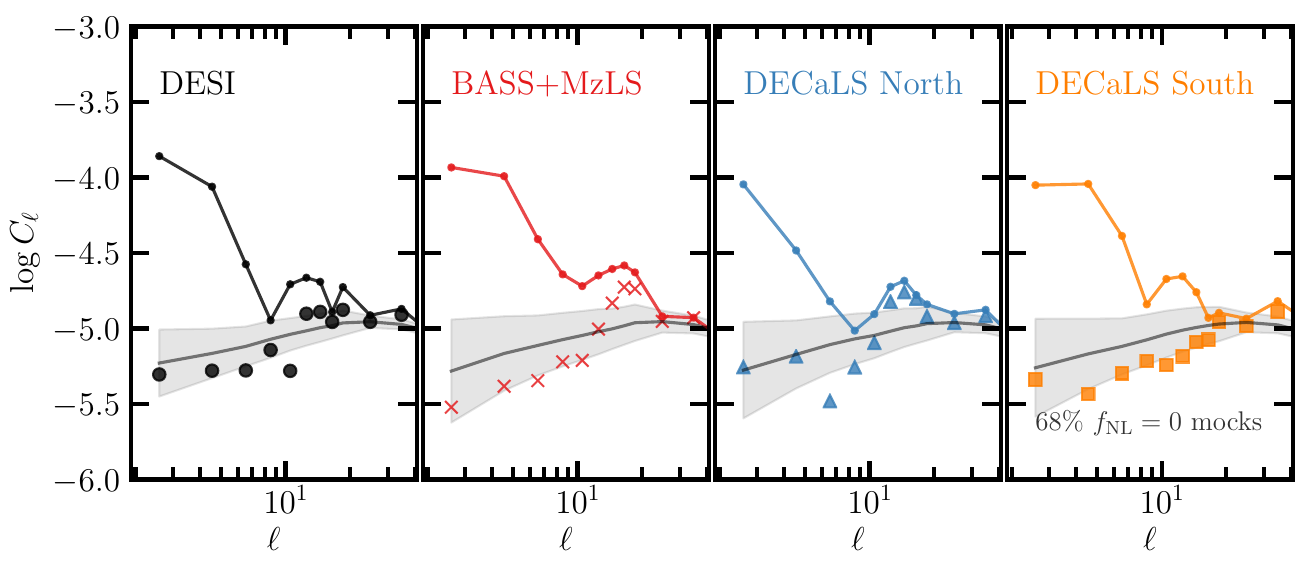}
    \includegraphics[width=0.9\textwidth]{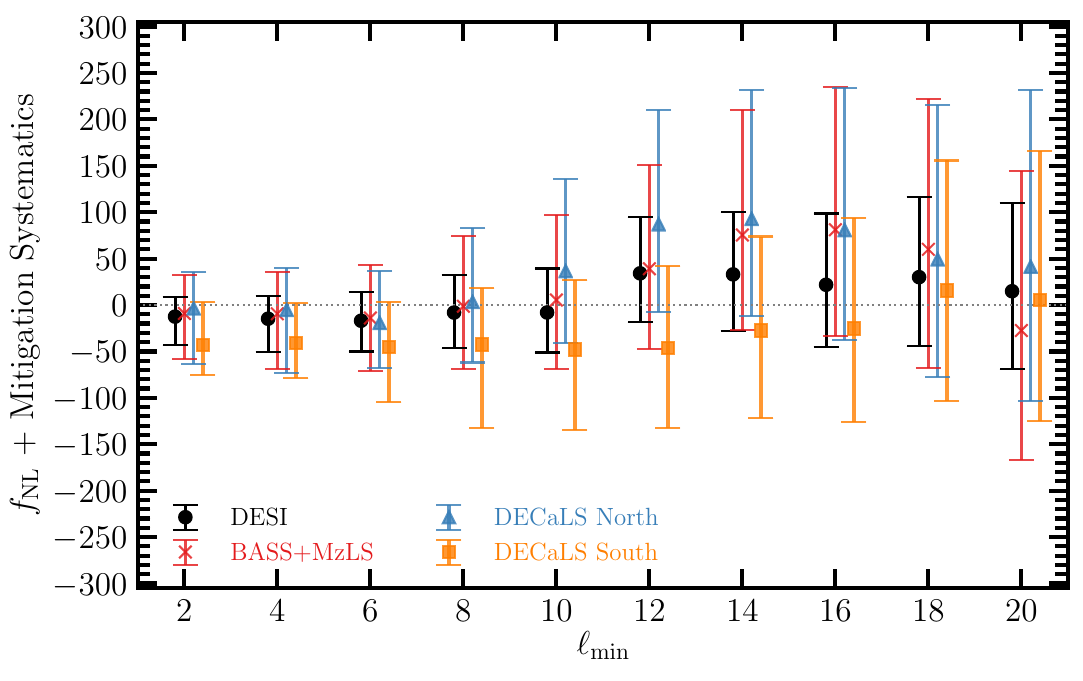}  
    \caption{Top: The measured power spectrum of the DESI LRG targets before (solid curves) and after \textit{non-linear nine maps} (scatter points) for the DESI, BASS+MzLS, DECaLS North, and DECaLS South regions. The solid curve and grey shade respectively represent the mean power spectrum and $68\%$ error from the $\fnl=0$ mocks with the same angular mask for each region. Bottom: The uncalibrated $\fnl$ constraints vs the lowest $\ell$ mode used for fitting $\fnl$. The points represent the best fitting estimates of $\fnl$ and error bars represent $95$\% confidence. The scaling of $\fnl$ is not calibrated to account for over-correction caused by mitigation.}\label{fig:mcmc_dr9elmin}
\end{figure*}

Now we proceed to perform some robustness tests and assess how sensitive the $\fnl$ constraints are to the assumptions made in the analysis or the quality cuts applied to the data. For each case, we re-train the cleaning methods and derive new sets of imaging weights. Accordingly, for the cases where a new survey mask is applied to the data, we re-calculate the covariance matrices using the new survey mask to account for the changes in the survey window and integral constraint effects. Calibrating the mitigation biases for all of these experiments is beyond the scope of this work and redundant, as we are only interested in the relative shift in the $\fnl$ constraints after changing the assumptions. Therefore, the absolute scaling of the $\fnl$ constraints presented here are biased because of the over-correction effect. Table \ref{tab:dr9method} summarizes the uncalibrated $\fnl$ constraints from the DESI LRG targets. Our tests are as follows:

\begin{itemize}[itemindent=*]

\item \textbf{Linear methods}: Even though the linear methods show remaining systematics (e.g., against depth in z as shown in Figure \ref{fig:nbarmock}), we obtain identical constraints from \textit{linear four maps} and \textit{linear nine maps}, respectively, $20(9)<\fnl<45(58)$ and $19(9)<\fnl<43(57)$ at $68\%(95\%)$ confidence. For the linear treatment methods, the probability of $\fnl$ being greater than zero is erroneously $99.9$ per cent. Any attempt to account for the over-correction would elevate this probability even further. The overestimation of $\fnl$ can be attributed to an increase in systematic contamination.
\item \textbf{Imaging regions}: We compare how our constraints from fitting the power spectrum of the whole DESI footprint compares to that from the power spectrum of each imaging region individually, namely BASS+MzLS, DECaLS North, and DECaLS South. Figure~\ref{fig:mcmc_dr9reg} shows the $68\%$ and $95\%$ probability contours on $\fnl$ and $b$ from each individual region, compared with that from DESI. The cleaning method here is \textit{non-linear nine maps}, and the covariance matrices are estimated from the $\fnl=0$ mocks. The bias in DECaLS North is lower than the ones from DECaLS South and BASS+MzLS, which might indicate some remaining systematic effects that could not be mitigated with the available imaging systematic maps. This is because given the negative correlation between $b$ and $\fnl$, a larger value of $\fnl$ due to excess clustering power needs to be compensated by a smaller value of $b$. Overall, we find that the constraints from analyzing each imaging survey separately are consistent with each other and DESI within $68\%$ confidence. We also consider combining the data at the likelihood level (`BASS+MzLS+DECaLS'). In this case the total number of data points is $111$ ($3\times 37$). We allow the bias and shotnoise paramters to vary independently for each sub-region but use a single and common $\fnl$ value, which brings the total number of free parameters to $7$ and the number of degrees of freedom to $104$. We obtain a best-fitting estimate of $\fnl=-31$ with $\chi^{2}=114.2$ and $68\%$ ($95\%$) confidence interval of $-41(-53) < \fnl < 9(5)$. Compared with our fiducial analysis which combines the data at the map level, we observe around $13\%$ loss in constraining power.

\item \textbf{Stellar density template (\textit{nStar})}: When not accounting for over-correction, adding the stellar density map appears to result in significant changes in the $\fnl$ constraints, e.g., compare non-linear three maps with non-linear four maps in Table \ref{tab:dr9method}. But these changes disappear when we account for the mitigation bias and we find both methods recover the same maximum likelihood estimate for $\fnl \sim 46$ within $69\%$ confidence, see Table \ref{tab:dr9methodcalib}, which implies that these changes can be associated with the over-correction issue from the chance correlations between the stellar density map and large-scale structure.

\item \textbf{Pixel completeness (\textit{comp. cut})}: We discard pixels with fractional completeness less than half to assess the effect of partially complete pixels on $\fnl$. This pixel completeness cut removes $0.6\%$ of the survey area, and no significant changes in the $\fnl$ constraints are observed.

\item \textbf{Imaging quality (\textit{imag. cut})}: Pixels with poor photometry are removed from our sample by applying the following cuts on imaging; $E[B-V]<0.1$, $nStar < 3000$, ${\rm depth}_{g} > 23.2$, ${\rm depth}_{r} > 22.6$, ${\rm depth}_{z} > 22.5$, ${\rm psfsize}_{g}<2.5$, ${\rm psfsize}_{r}<2.5$, and ${\rm psfsize}_{z}<2$. Although these cuts remove $8\%$ of the survey mask, there is a negligible impact on the best-fitting estimates of $\fnl$ from fitting the DESI power spectrum. However, when each region is fit individually, the BASS+MzLS constraint is more stable than those from DECaLS North and DECaLS South. 

\item \textbf{Covariance matrix (\textit{cov})}: We fit the power spectrum of our sample cleaned with \textit{non-linear nine maps}, but use the covariance matrix constructed from the $\fnl=76.9$ mocks. With the alternative covariance, a $7\%$ increase in the 68\% error on $\fnl$, $\sigma(\fnl)$, is observed. We also find that the best-fitting and marginalized mean estimates of $\fnl$ increase slightly by $\Delta \fnl = 2$. Overall, we find that the differences are not significant in comparison to the statistical precision.

\item \textbf{External maps (\textit{CALIBZ+HI})}: The neural network eleven maps correction includes the additional maps for the neutral column density (HI) and the z-band calibration error (CALIBZ). With this correction, the best-fitting $\fnl$ increases from $-4$ to $-2$ for DECaLS North and from $-43$ to $-38$ for DECaLS South, which might suggest that adding HI and CALIBZ increases the input noise, and thus negatively impacts the performance of the neural network model. This test is not performed on BASS+MzLS due to a lack of coverage from the CALIBZ map. 

\item \textbf{Declination mask (\textit{no DEC cut})}: The fiducial mask removes the disconnected islands in DECaLS North and regions with DEC $<-30$ in DECaLS South, where there is a high likelihood of calibration issues as different standard stars are used for photometric calibrations. We analyze our sample without these cuts, and find that the best-fitting and marginalized $\fnl$ mean estimates from DECaLS South shift significantly to higher values of $\fnl$ by $\Delta \fnl \sim 41$, which supports the case that there are remaining photometric systematics in the DECaLS South region below DEC $=-30$. On the other hand, the constraints from DECaLS North do not change significantly, indicating the islands do not induce significant contaminations.

\item \textbf{Scale dependence (\textit{varying $\ell_{\rm min}$})}: We raise the value of the lowest harmonic mode $\ell_{\rm min}$ used for the likelihood evaluation during MCMC. This is equivalent to utilizing smaller spatial scales in the measurements of the power spectrum. By doing so, we anticipate a reduction in the impact of imaging systematics on $\fnl$ inference as lower $\ell$ modes are more likely to be contaminated. Figure \ref{fig:mcmc_dr9elmin} illustrates the power spectra before and after the correction with \textit{non-linear nine maps} in the top panel. The bottom panel shows the best fitting estimate and $95\%$ error on $\fnl$ with \textit{non-linear nine maps} for the DESI, BASS+MzLS, DECaLS North, and DECaLS South regions. We discover that a slight upward shift in the best fitting estimates of $\fnl$ on scales ranging from $10$ to $20$ for DECaLS North and BASS+MzLS when we utilized a higher $\ell_{\rm min}$. This outcome might imply that the imaging systematic maps do not contain enough information to help the cleaning method null out the contaminating signal in the NGC. We also find that the bump is resilient against an alternative correction, in which we apply the neural networks trained on the DECaLS South to the DECaLS North region (see \ref{ssec:ndecalsbump}). Overall, this result is contrary to what one might predict if a significant systematic-induced spike existed at the very low $\ell$, or if we had an extremely large-scale systematic leakage from the $\ell=1$ mode. As a result, it suggests that the underlying issue is more subtle than originally anticipated.
\end{itemize}

%% file: sections/conclusion.tex
\section{Discussion and Conclusion}\label{sec:conclusion}

We have measured the local PNG parameter $\fnl$ using the scale-dependent bias in the angular clustering of LRGs selected from the DESI Legacy Imaging Survey DR9. Our sample includes more than $12$ million LRG targets covering around $14,000$ square degrees in the redshift range of $0.2< z < 1.35$. We leverage early spectroscopy during DESI Survey Validation \citep{desi2023sv} to infer the redshift distribution of our sample (Figure \ref{fig:nz}). Our power spectrum model accounts for various theoretical and observational effects such as RSD, magnification bias, survey geometry, and integral constraint. Most importantly, we utilize a novel machine learning-method to mitigate the effect of imaging systematics and reduce excess clustering power on large scales (or low $\ell$). We use lognormal simulations to estimate the covariance matrices. As a caveat, this omits the contributions from higher order statistics in the covariance matrix, but we leave that for future as we do not anticipate any major impact on the best fitting estimates of $\fnl$.

\begin{figure}
    \centering
    \includegraphics[width=0.45\textwidth]{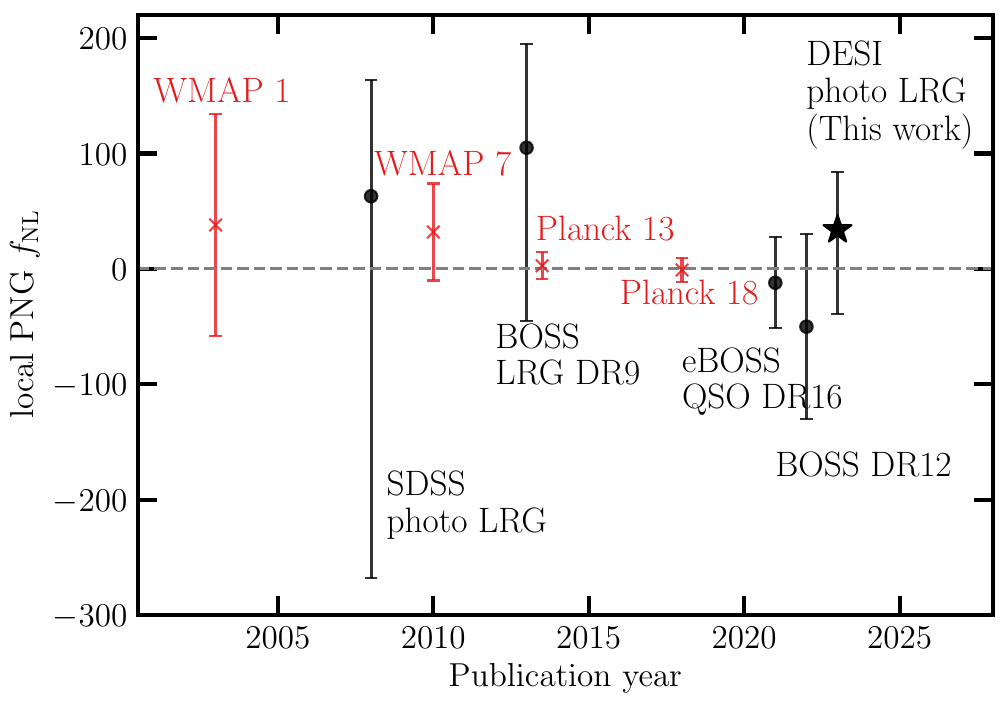}
    \caption{History of constraints on local PNG $\fnl$ at $95\%$ confidence from single-tracer LSS \citep{slosar2008constraints,2013MNRAS.428.1116R, mueller2022primordial, 2022PhRvD.106d3506C}, including our analysis with $-39<\fnl<84$ (DESI photo LRG) and CMB surveys \citep{Komatsu_2003, Komatsu_2010, planck13, akrami2019planck}. The median $\fnl$ value is used in case the maximum likelihood estimate was not reported in the reference.}
    \label{fig:fnlhist}
\end{figure}

In our fiducial analysis, which includes a non-linear treatment of systematics using nine imaging property maps (Galactic extinction, stellar density, depth in $grzW1$, and psfsize in $grz$), we obtain $\fnl = 34^{+24(+50)}_{-44(-73)}$ with $p=1$ and $s=0.945$. This measurement is consistent with recent CMB and LSS measurements, as visualized in Figure \ref{fig:fnlhist}. The sensitivity of our constraints is explored against $p$ and $s$. The best fitting estimates of $\fnl$ decrease as we increase either $p$ or $s$. Specifically, we find that the error on $\fnl$ is more sensitive to $p$ than $s$. Compared with the fiducial result, the error increases by more than a factor of two for $p=1.6$, and only by $7\%$ for $s=1.25$ (Figure \ref{fig:fnl_magbias}). The minimum $\chi^{2}$ however does not change much, indicating that the impact on the power spectrum fit is negligible.

The signature of local PNG is very sensitive to excess clustering power caused by imaging systematic effects. We have applied a series of robustness tests to investigate the impact of how the galaxy selection function is determined. Specifically, both linear and nonlinear methods are applied using various combinations of imaging systematic maps (including two external maps for the neutral hydrogen column density and photometric calibration error in the z band). We also examine the effect of additional masks based on imaging conditions and survey completeness. Overall, we find that no change in the analysis shifts the maximum likelihood value of $\fnl$ to a significantly different value (Figure \ref{fig:mcmc_dr9reg}, Figure \ref{fig:mcmc_dr9elmin}, and Table \ref{tab:dr9method}).

Although being essential for the mitigation of imaging systematics, the template-based approach inevitably removes some of the large-scale clustering information. One of the primary highlights of this work is that we present a strategy to calibrate the systematic mitigation's impact on the inferred $\fnl$. As we increase the number of maps for mitigation, more of the power spectrum is removed, introducing a larger bias to the $\fnl$ posterior distribution. Our mock tests suggest that this bias is $\fnl$-dependent, such that the mocks with larger $\fnl$ experience a more substantial reduction in the low$-\ell$ power due to systematic mitigation. Therefore, it is crucial to calibrate for this effect using simulations that have gone through the same treatment methods and are subject to the same over-correction effect.

As a greater flexibility in the mitigation increases the over-correction and decreases the statistical power, we tested if we can reduce the flexibility in our method by using a smaller set of maps, including Galactic extinction, depth in the z band, and astronomical seeing in the r band (nonlinear three maps) to retain some constraining power. Additionally, we consider an additional map for local stellar density (nonlinear four maps). Using three or four maps, we can qualitatively mitigate systematic trends in the mean galaxy density and cross correlations of the galaxy density field and imaging property maps (see Section \ref{sec:systests}). These methods do not degrade the error on $\fnl$ as much as the fiducial method which used nine maps. However, when applying our null-tests that are applied in order to detect residual systematic variance (see \ref{sec:systests}), we obtain passing results only for the nonlinear nine map case. In this work, we found updating the covariance matrix for each particular variation (e.g., the mitigation method applied) was important in order to obtain a similar $\chi^2$ of the null test when applied to the mocks and hence self-consistently obtain a $p$-value for the null test. Another important conclusion from applying our null tests to mocks is that the mean density contrast test is less sensitive to the $\fnl$ value for the mocks than the angular cross-power. Given the amount of $\fnl$ constraining power that we lose when applying the nine map regression (the uncertainties approximately double), our findings highlight the importance of exploring, developing, and validating alternative mitigation approaches to avoid over-correction for a robust analysis of local PNG.

Our analysis can be considered as the first attempt to identify major systematics in DESI, so we can be ready for constraining $\fnl$ with DESI spectroscopy. Internal DESI tests of the photometric calibration were unable to uncover DESI-specific issues, e.g., when comparing to Gaia data. The most significant trends that we find are with the E(B-V) map. The source of such a trend would be a mis-calibration of the E(B-V) map itself or the coefficients applied to obtain Galactic extinction corrected photometry. Such a mis-calibration would plausibly be proportional in amplitude to the estimated E(B-V) map, though it may not have E(B-V)’s spatial distribution. There are ongoing efforts within DESI to obtain improved Galactic extinction information, which will help us address systematics. Additionally, cross-correlations of the DESI LRG density with the CMB lensing map is more stable in terms of systematics and can complement the results presented in this work. We can further avoid the over-fitting issue by combining our neural network-based treatment method with forward-modeling techniques, such as Obiwon \citep{kong2020}, but we will leave that for future work.

%% file: sections/acknowledgement.tex
\section*{Acknowledgements}
We would like to thank Douglas Finkbeiner for feedback on an early version of the manuscript; Violeta Gonzalez-Perez for handling the DESI internal review process; Tanveer Karim, Sukhdeep Singh, Ahmad Shamloumehr, and Reza Katebi for helpful discussions; and Rongpu Zhou for estimating the slope of the number counts and providing the maps for galaxy density and imaging systematics. MR would like to thank Ohio State's Center for Cosmology and AstroParticle Physics, in particular, John Beacom and Lisa Colarosa, for their hospitality and support. MR is supported by the U.S. Department of Energy grants DE-SC0021165 and DE-SC0011840. H-JS acknowledges support from the U.S. Department of Energy, Office of Science, 
Office of High Energy Physics under grant No. DE-SC0019091 and No. DE-SC0023241. AP acknowledges support from the UK Science and Technology Facilities Council (STFC) under grant numbers ST/V000594/1 and from the European Union’s Horizon Europe program under the Marie Skłodowska-Curie grant agreement 101068581. FB is a University Research Fellow, and has received funding from the European Research Council (ERC) under the European Union’s Horizon 2020 research and innovation program (grant agreement 853291). BB-K is supported by the project \begin{CJK}{UTF8}{mj}우주거대구조를 이용한 암흑우주 연구\end{CJK} (``Understanding Dark Universe Using Large Scale Structure of the Universe’’), funded by the Ministry of Science of the Republic of Korea. We acknowledge the support and resources from the Ohio Supercomputer Center \citep[OSC;][]{OhioSupercomputerCenter1987}. This research has made substantial use of the arXiv preprint server, NASA’s Astrophysics Data System, Github's online software development platform, and many open-source software, such as Pytorch, Nbodykit, HEALPix, Fitsio, Scikit-Learn, NumPy, SciPy, Pandas, IPython, and Jupyter.

This material is based upon work supported by the U.S. Department of Energy (DOE), Office of Science, Office of High-Energy Physics, under Contract No. DE–AC02–05CH11231, and by the National Energy Research Scientific Computing Center, a DOE Office of Science User Facility under the same contract. Additional support for DESI was provided by the U.S. National Science Foundation (NSF), Division of Astronomical Sciences under Contract No. AST-0950945 to the NSF's National Optical-Infrared Astronomy Research Laboratory; the Science and Technology Facilities Council of the United Kingdom; the Gordon and Betty Moore Foundation; the Heising-Simons Foundation; the French Alternative Energies and Atomic Energy Commission (CEA); the National Council of Science and Technology of Mexico (CONACYT); the Ministry of Science and Innovation of Spain (MICINN), and by the DESI Member Institutions: \href{https://www.desi.lbl.gov/collaborating-institutions}{https://www.desi.lbl.gov/collaborating-institutions}.

The DESI Legacy Imaging Surveys consist of three individual and complementary projects: the Dark Energy Camera Legacy Survey (DECaLS), the Beijing-Arizona Sky Survey (BASS), and the Mayall z-band Legacy Survey (MzLS). DECaLS, BASS and MzLS together include data obtained, respectively, at the Blanco telescope, Cerro Tololo Inter-American Observatory, NSF's NOIRLab; the Bok telescope, Steward Observatory, University of Arizona; and the Mayall telescope, Kitt Peak National Observatory, NOIRLab. NOIRLab is operated by the Association of Universities for Research in Astronomy (AURA) under a cooperative agreement with the National Science Foundation. Pipeline processing and analyses of the data were supported by NOIRLab and the Lawrence Berkeley National Laboratory. Legacy Surveys also uses data products from the Near-Earth Object Wide-field Infrared Survey Explorer (NEOWISE), a project of the Jet Propulsion Laboratory/California Institute of Technology, funded by the National Aeronautics and Space Administration. Legacy Surveys was supported by: the Director, Office of Science, Office of High Energy Physics of the U.S. Department of Energy; the National Energy Research Scientific Computing Center, a DOE Office of Science User Facility; the U.S. National Science Foundation, Division of Astronomical Sciences; the National Astronomical Observatories of China, the Chinese Academy of Sciences and the Chinese National Natural Science Foundation. LBNL is managed by the Regents of the University of California under contract to the U.S. Department of Energy. The complete acknowledgments can be found at \href{https://www.legacysurvey.org/}{https://www.legacysurvey.org/}.

Any opinions, findings, and conclusions or recommendations expressed in this material are those of the author(s) and do not necessarily reflect the views of the U. S. National Science Foundation, the U. S. Department of Energy, or any of the listed funding agencies.

The authors are honored to be permitted to conduct scientific research on Iolkam Du’ag (Kitt Peak), a mountain with particular significance to the Tohono O’odham Nation.

%% file: sections/dataavailability.tex
\section*{Data Availability}
\label{sec:dataavail}
The DR9 catalogues from the DESI Legacy Imaging Surveys are publicly available at \href{https://www.legacysurvey.org/dr9/}{https://www.legacysurvey.org/dr9/}. The software used for cleaning the imaging data is available at \href{https://github.com/mehdirezaie/sysnetdev}{https://github.com/mehdirezaie/sysnetdev}. All data points shown in the published graphs are available in a machine-readable form at \href{https://zenodo.org/records/10594656}{https://zenodo.org/records/10594656}.

%% file: sections/extra.tex
\section{Extra robustness tests}

\subsection{Scale dependent systematics}\label{sec:scalesys}
\begin{figure}
\centering
\includegraphics[width=0.5\textwidth]{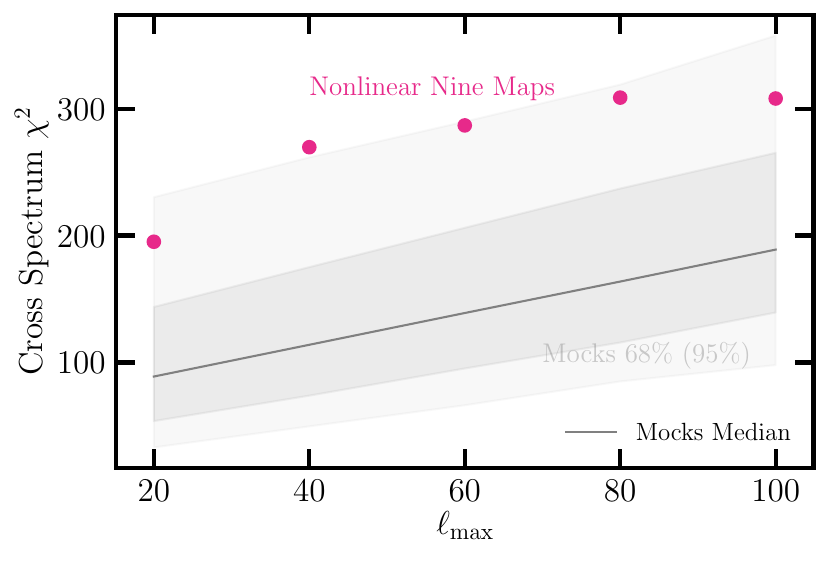}
\caption{The cross power spectrum's $\chi^{2}$ between the DESI LRG target density and imaging systematic maps as a function of the highest mode $\ell_{\rm max}$ when the sample is cleaned with the linear (triangles) and non-linear (squares) three maps. The lowest mode is fixed at $\ell_{\rm min}=2$. The solid curve and dark (light) shade represent the median value and $68\%$ ($95\%$) confidence regions, estimated from the $\fnl=0$ mocks.}\label{fig:chi2cellextend}
\end{figure}

To investigate the statistical significance of the cross power spectrum's $\chi^{2}$, we examine its dependence on the largest harmonic mode $\ell_{\rm max}$. Our fiducial cross power spectrum diagnostic (equation \ref{eq:cx_chi2}) uses harmonic modes up to $\ell=20$, which determines the smallest scale used for characterizing residual systematic errors. We extend $\ell_{\rm max}$ from $20$ to $100$, where the latter scale corresponds to density fluctuations on scales smaller than $2$ degrees. Figure \ref{fig:chi2cellextend} shows the median of the normalized cross power spectrum's $\chi^{2}$ from the clean $\fnl=0$ mocks after non-linear nine maps as the highest mode $\ell_{\rm max}$ increases from $20$ to $100$ (represented by the solid line). The pink circles represent the $\chi^{2}$ values for the DESI LRG targets cleaned with the non-linear nine maps method. Overall, we find that for all scales up to $\ell=100$, the nonlinear nine maps approach yields consistent values with the clean mocks.

\subsection{Survey window convolution}\label{ssec:windowconv}
\begin{figure}
    \centering
    \includegraphics[width=0.5\textwidth]{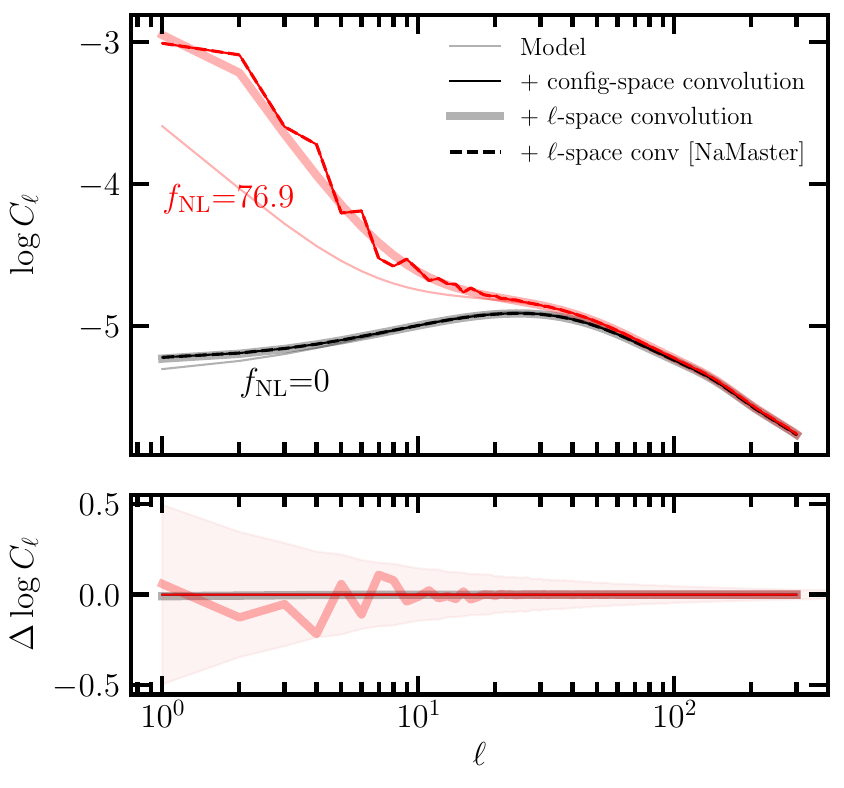}
    \caption{The model power spectrum before and after the survey geometry convolution for $\fnl=0$ and $76.9$ using the DESI survey mask. The bottom panel shows the residual error with respect to the \textsc{NaMaster} code. The shade represents the dispersion of the $\fnl=76.9$ mocks.}
    \label{fig:window_conv}
\end{figure}

Here we calculate the mode-mode coupling matrix from the DESI mask. This matrix depends only on the survey geometry and can be described in terms of the window power spectrum \citep{hivon2002master},
\begin{equation}
M_{\ell \ell^{\prime}} = \frac{2\ell^{\prime}+1}{4\pi} \sum_{\ell^{\prime\prime}} (2\ell^{\prime\prime}+1) \tilde{C}^{\rm window}_{\ell^{\prime\prime}} \begin{pmatrix}
\ell & \ell^{\prime} & \ell^{\prime\prime}\\
0 & 0 & 0
\end{pmatrix},
\end{equation}
where the last term in the right hand side represents the Wigner 3-j symbol (or Clebsch-Gordan coefficient), and is calculated using \textsc{SymPy} \citep{sympy2017}. We benchmark our code against the publicly available software, \textsc{NaMaster}\footnote{\href{https://github.com/LSSTDESC/NaMaster}{https://github.com/LSSTDESC/NaMaster}} \citep{2019MNRAS.484.4127A}. Figure \ref{fig:window_conv} illustrates various approaches to address the mode-mode coupling resulting from the DESI survey window at two arbitrary values of $\fnl$. The red shade represents the 68\% dispersion of the $\fnl=76.9$ mocks. When $\fnl=0$, our \textit{config-space} convolution of the window aligns with the \textit{$\ell$-space} convolution approach. However, when $\fnl=76.9$, both the config-space and \textsc{NaMaster} $\ell$-space methods yield a convoluted power spectrum with noticeable noise-like numerical artifacts on large scales, therefore possibly in a $\fnl$-dependent manner. To assess the impact of these discrepancies on our $\fnl$ constraints, we fit the clustering of the DESI LRG targets, disregarding the integral constraint effect. The best-fitting estimates of $\fnl$ will be biased, but our focus is on understanding the relative impact on $\fnl$ between the two approaches. For both config-space and $\ell$-space methods, we obtain a similar minimum $\chi^{2}$ value of $39.6$ with $34$ degrees of freedom. Notably, the posterior width for the config-space approach is slightly larger than that of the $\ell$-space by $10\%$. The absolute difference in the best-fitting estimates of $\fnl$ between the two cases is less than $1.1$, considered negligible relative to the statistical precision of our measurements.

\subsection{Redshift uncertainties}\label{ssec:nzuncer}

\begin{figure}
\raggedleft
\includegraphics[width=0.46\textwidth]{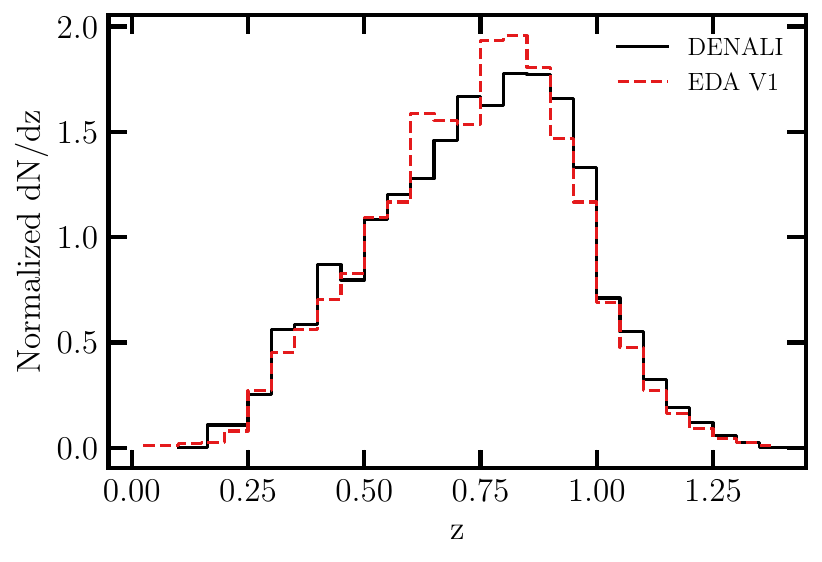}
\includegraphics[width=0.48\textwidth]{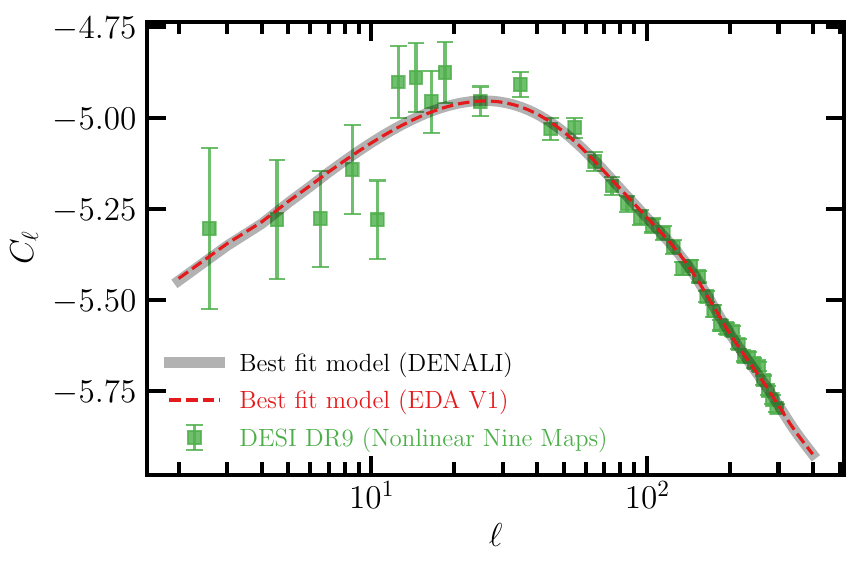}\caption{Top: The redshift distribution of the DESI LRG targets from the EDA V1 and Denali. Bottom: The measured power spectrum of the DESI LRG targets and the best fit theory models using different redshift distributions.}\label{fig:cl_nz}
\end{figure}

We use the Early Data Assembly Version 1 (EDA V1) to construct the redshift distribution for the DESI LRG targets. We find that the change in the maximum likelihood estimate of $\fnl$ is negligible, $|\Delta \fnl | < 1$, compared to the statistical precision of our measurements. Figure \ref{fig:cl_nz} shows the measured power spectrum of the DESI targets and the corresponding best fit theory curves. The variations in $dN/dz$ do not significicantly alter the conclusion of our paper.

\subsection{Spurious bump in NGC}\label{ssec:ndecalsbump}

\begin{figure}
    \centering
    \includegraphics[width=0.48\textwidth]{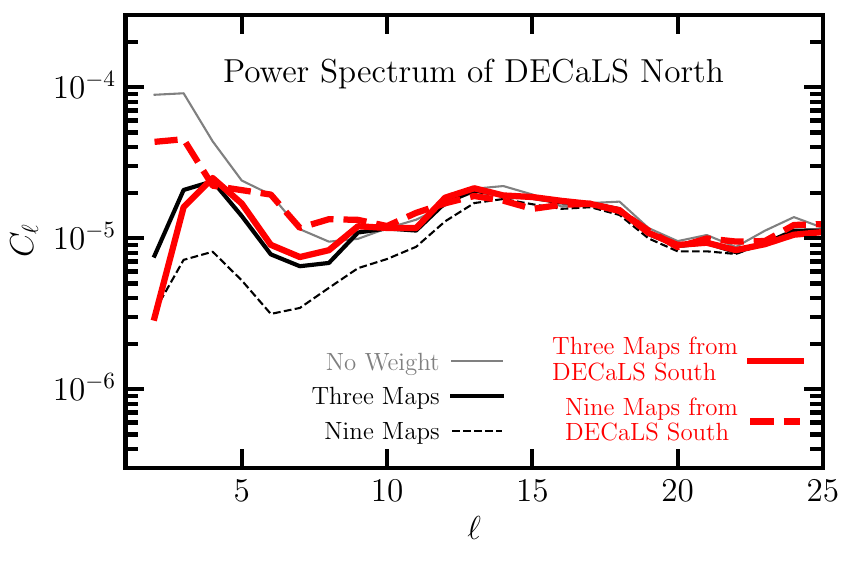}
    \caption{The unbinned measured power spectrum of the DESI LRG targets in the DECaLS North region before and after various mitigations using the neural network approach.}
    \label{fig:clSonN}
\end{figure}

As shown in Figure \ref{fig:mcmc_dr9elmin} (top panel), we realize that the spurious feature at $\ell \sim 10-20$ is removed in the DECaLS South region after mitigation, but it remains in the BASS+MzLS and DECaLS North. We use the neural networks trained on the DECaLS South with three and nine maps to mitigate the galaxy density in the DECaLS North region, and then measure the power spectrum. Figure \ref{fig:clSonN} shows the power spectrum before treatment (No Weight) and after the nonlinear three maps and nine maps methods for comparison. We find that whatever causing the bump is different between the DECaLS North and South. The best-fit estimates for $\fnl$ from the DR9 DECaLS North using the neural network correction with three maps (NN trained on DECaLS North), three maps (NN trained on DECaLS South), and nine maps (NN trained on DECaLS South) are $41$, $36$, and $75$, respectively. The solution without correction (No weight) results in a best fitting estimate of $\fnl=94$.

\section{Lognormal mocks}

We fit the mean power spectrum of the lognormal mocks to validate the modeling pipeline, and in particular the survey geometry and integral constraint treatments. We investigate the impact of covariance matrix on the inference of $\fnl$. Finally, we show the impact of imaging systematic mitigation and the over-subtraction effect when the cleaning methods are applied to the mocks. 

\subsection{Clean mocks}
\begin{figure}
    \centering
    \includegraphics[width=0.48\textwidth]{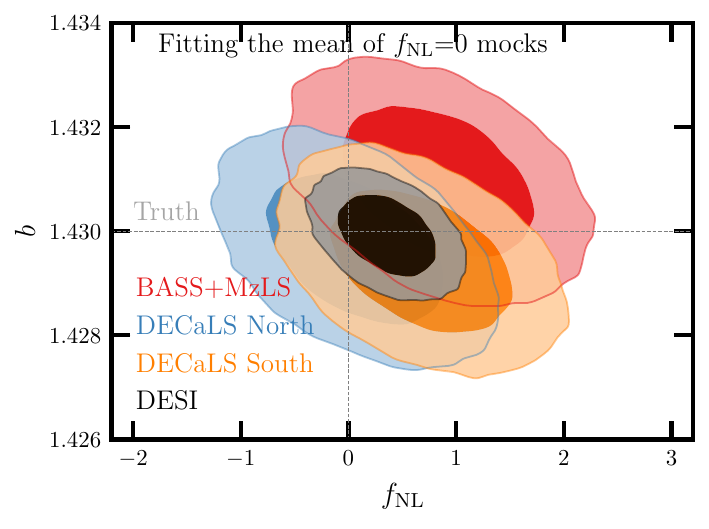} 
    \caption{68\% and 95\% confidence contours from the mean power spectrum of the $\fnl=0$ mocks for the DESI footprint and sub-imaging surveys. The truth values are represented by vertical and horizontal lines.}\label{fig:mcmc_mocks0}
\end{figure}

\begin{figure}
    \centering
    \includegraphics[width=0.48\textwidth]{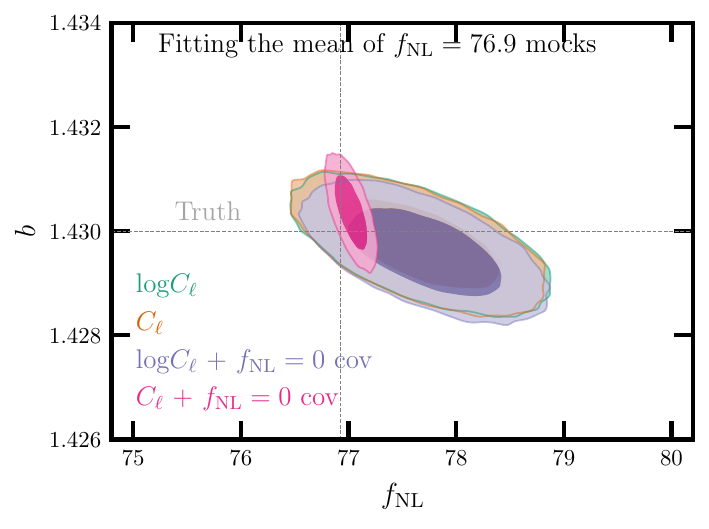} 
    \caption{68\% and 95\% confidence contours of fitting the mean power spectrum or its log transformation from the $\fnl=76.9$ mocks for the DESI footprint. Using the $\log C_{\ell}$ fitting yield constraints that are insensitive to the covariance used. The truth values are represented by vertical and horizontal lines.}\label{fig:mcmc_mocks100}
\end{figure}

The $68\%$ and $95\%$ probability contours on the PNG parameter $\fnl$ and bias coefficient $b$ are shown in Figure \ref{fig:mcmc_mocks0} and \ref{fig:mcmc_mocks100}, respectively, for the $\fnl=0$ and 76.9 mocks. The best-fitting, marginalized mean estimates, as well as the $1\sigma$ and $2\sigma$ confidence intervals of $\fnl$ are summarized in Table \ref{tab:mocksmcmc}. 

Measuring the power spectrum from the entire DESI footprint reduces the cosmic variance and thus improves the constraining power. Figure \ref{fig:mcmc_mocks0} compares the constraints from fitting the log of the mean power spectrum of the mocks when it is measured from the DESI footprint to those obtained from the sub imaging surveys. We find that the underlying true $\fnl$ value is recovered within $95\%$ confidence, and that the contours for the DESI region are smaller by a factor of two. 

\begin{figure}
    \centering
    \includegraphics[width=0.45\textwidth]{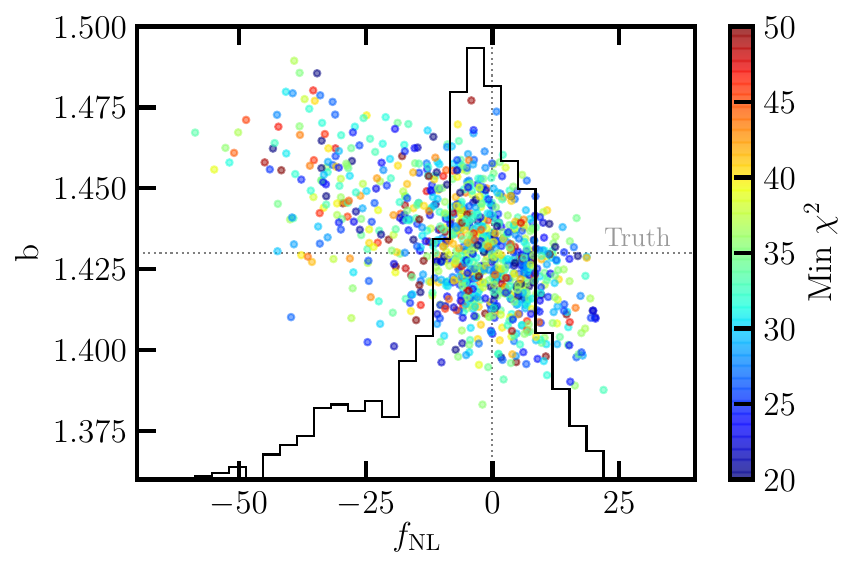} 
    \includegraphics[width=0.45\textwidth]{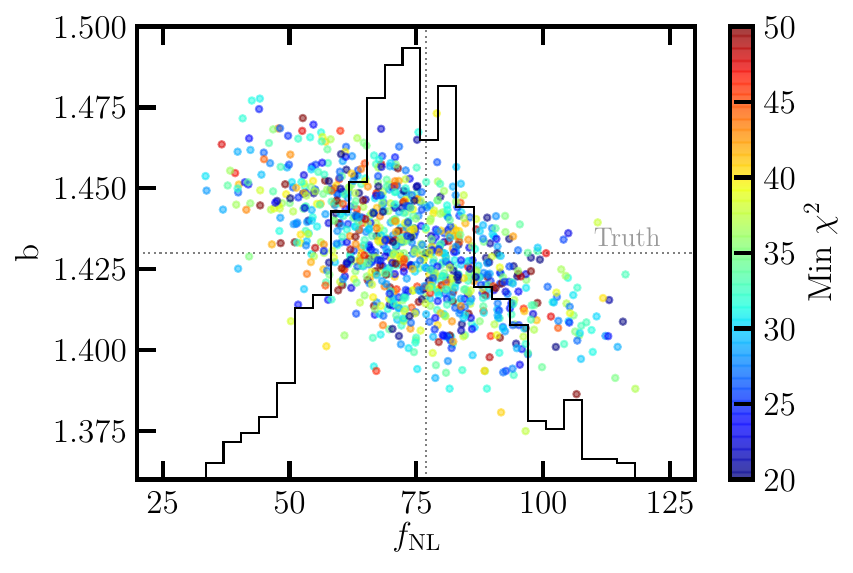}         
    \caption{The best-fitting estimates of $b$ and $\fnl$ from fitting 1000 lognormal mocks with $\fnl=0$ (top) and $76.9$ (bottom) in the DESI footprint. No mitigation is applied to the mocks. The truth values are represented by vertical and horizontal lines.}\label{fig:bestfit_mocks}
\end{figure}

The power spectrum of the mocks at low $\ell$ is very sensitive to the cosmic variance and the true value of $\fnl$. Consequently, a large value of $\fnl$ can induce very large power on low $\ell$, and thus significantly change the covariance matrix. We find that applying the log transformation on the power spectrum makes the result more robust against the choice of the covariance matrix. Figure \ref{fig:mcmc_mocks100} shows the confidence contours when we fit either the power spectrum or its log transform of the $\fnl=76.9$ mocks, and use different covariance matrices. We consider the $\fnl=0$ and $76.9$ mocks to construct the covariance from one set and use it to fit the mean power spectrum of the other set. When the covariance matrix is constructed from the same set of mocks used for the mean power spectrum, we find that the difference in $\fnl$ constraints between fitting the power spectrum and its log transformation is negligible at only 2\%. If we use the $\fnl=0$ mocks to estimate the covariance, and fit the log power spectrum of the $\fnl=76.9$ mocks, we find that the error on $\fnl$ increases only by $7\%$. However, when the mean power spectrum of the $\fnl=76.9$ mocks is fit using the covariance matrix estimated from the $\fnl=0$ mocks, the constraints tighten by a factor of $5$ due to a higher signal to noise ratio. Therefore, we argue that fitting the log power spectrum can help mitigate the need for having $\fnl$-dependent covariance matrices and make the constraints less sensitive to covariance construction.

\begin{table*}
    \caption{The best-fitting and marginalized mean estimates for $\fnl$ from fitting the mean power spectrum of the mocks. The covariance is scaled to represent the error on the mean power spectrum. The number of degrees of freedom is 34 (37 data points - 3 parameters).}   
    \label{tab:mocksmcmc}  
   \centerline{%
    \begin{tabular}{llllllll}
    \hline
    \hline
   &  & & & & $\fnl$ &  \\
   \cmidrule(r{.7cm}){4-7}    
Mock / $\fnl$ &  Footprint   &  Observable & 	Best fit  & Mean & $ 68\%$ CL & $ 95\%$ CL & $\chi^{2}$ (dof $=34$)\\
    \hline
Clean $76.9$ & DESI & log$C_{\ell}$    & $ 77.67$& $ 77.67$& $ 77.17<\fnl< 78.16$& $ 76.71<\fnl< 78.64$ &   38.8\\
Clean $76.9$ & DESI & $C_{\ell}$       & $ 77.67$& $ 77.65$& $ 77.17<\fnl< 78.14$& $ 76.70<\fnl< 78.60$ &   39.0\\
Clean $76.9$ & DESI & log$C_{\ell}$ + $f_{\rm NL}=0$ cov & $ 77.70$& $ 77.71$& $ 77.25<\fnl< 78.17$& $ 76.81<\fnl< 78.63$ &   39.9\\
Clean $76.9$ & DESI & $C_{\ell}$ + $f_{\rm NL}=0$ cov & $ 77.03$& $ 77.02$& $ 76.93<\fnl< 77.12$& $ 76.83<\fnl< 77.22$ &  207.6\\
\hline
Clean $0$ & DESI         &  log$C_{\ell}$ & $  0.36$& $  0.36$& $  0.06<\fnl<  0.65$& $ -0.23<\fnl<  0.94$ &   35.7\\
Clean $0$ & BASS+MzLS    &  log$C_{\ell}$ & $  0.83$& $  0.82$& $  0.25<\fnl<  1.40$& $ -0.31<\fnl<  1.96$ &   39.4\\
Clean $0$ & DECaLS North &  log$C_{\ell}$& $  0.07$& $  0.06$& $ -0.47<\fnl<  0.60$& $ -1.00<\fnl<  1.12$ &   26.7\\
Clean $0$ & DECaLS South &  log$C_{\ell}$& $  0.67$& $  0.67$& $  0.13<\fnl<  1.22$& $ -0.40<\fnl<  1.75$ &   34.3\\
\hline
    \end{tabular}
    }
\end{table*}

Figure \ref{fig:bestfit_mocks} shows the best-fitting estimates for $b$ vs $\fnl$ for $\fnl=0$ and $=76.9$ mocks in the top and bottom panels, respectively. Truth values are represented via the dotted lines. The points are color-coded with the minimum $\chi^{2}$ from fit for each realization. The histograms of the best-fitting $\fnl$ estimates are plotted in the background. For the $\fnl=0$ mocks, the best-fitting estimates are more symmetric. To understand this behaviour, we consider the first derivative of the likelihood (Equation \ref{eq:likelihood}), which is proportional to the first derivative of the log power spectrum. By simplifying the integrals involved in $C_{\ell}$, we have $C_{\ell} = A_{0, \ell} + A_{1,\ell} \fnl + A_{2, \ell} \fnl^{2}$ where $A_{123,\ell}$ are $\ell$-dependent terms. Then, the derivative of the likelihood is proportional to
\begin{equation}
    \frac{d}{d\fnl}\log(C_{\ell}) = \frac{A_{1, \ell}+2A_{2, \ell}\fnl}{A_{0, \ell} + A_{1,\ell} \fnl + A_{2, \ell} \fnl^{2}}.
\end{equation}
For infinitesimal values of $\fnl$, the derivative becomes asymptotically independent from $\fnl$ while for large values of $\fnl$ it decreases as $2/\fnl$. This implies that for the $\fnl=0$ mocks, the likelihood is more likely to be skewed toward negative values.

\subsection{Contaminated mocks}\label{ssec:contmocks}
\begin{figure*}
    \centering
    \includegraphics[width=0.9\textwidth]{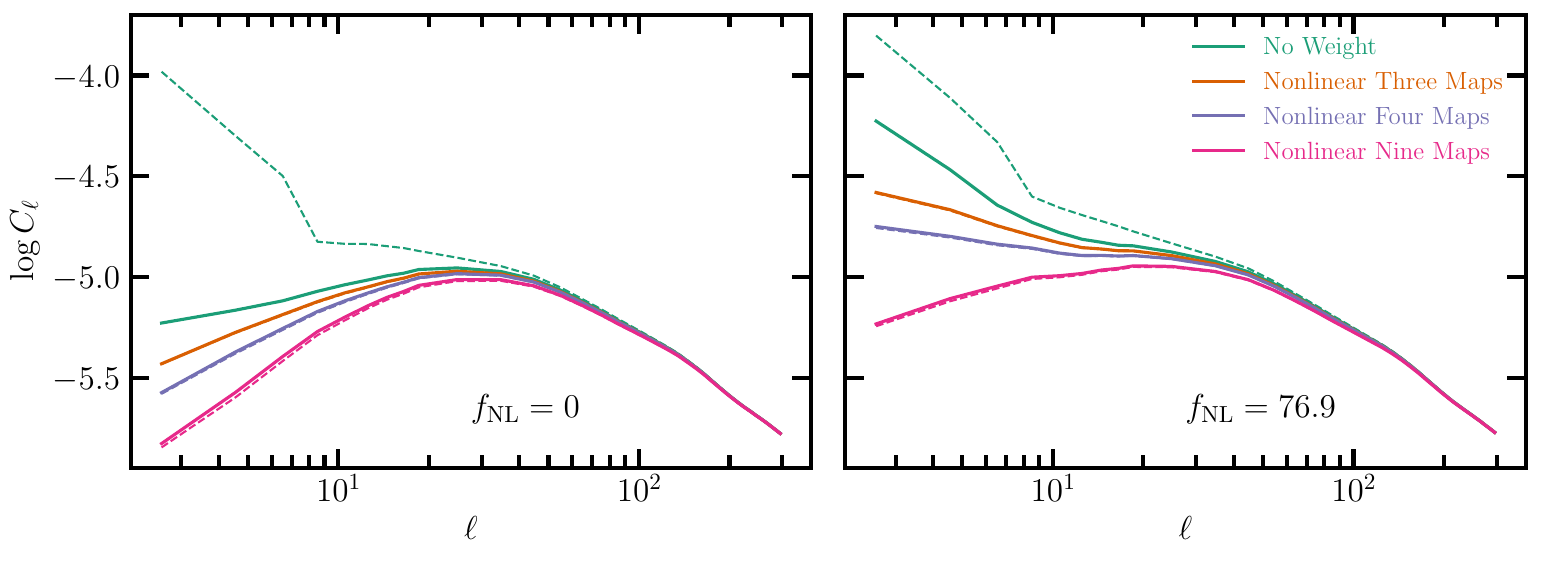}
    \caption{The mean power spectrum of the $\fnl =0$ and $76.9$ mocks with (dashed) and without (solid) imaging systematics before ('No Weight') and after applying the non-linear cleaning method with three, four, and nine maps.}\label{fig:clmocks}
\end{figure*}

Our nonlinear neural network-based approach is applied to the $\fnl=0$ and $76.9$ mocks. We only consider the methods that include running the neural network with three, four, and nine imaging systematic maps. The measured mean power spectrum of the mocks are shown in Figure \ref{fig:clmocks} for $\fnl=0$ (left) and $76.9$ (right). The solid and dashed curves show the measurements respectively from the clean and contaminated mocks.

We find that the imaging treatment has removed some of the true clustering signal, and the amount of the over-subtraction is almost the same regardless of whether the mocks have systematics. The over-subtraction induces biases in the $\fnl$ constraints, as summarized in Table \ref{tab:contmocksmcmc}. The over-subtraction at low $\ell$ is so high that we get a poor fit after applying the mitigation with the nonlinear three maps approach, e.g., $\chi^{2}=86.8$ for the clean $\fnl=0$ mocks.

\begin{figure*}
\centering
\includegraphics[width=0.45\textwidth]{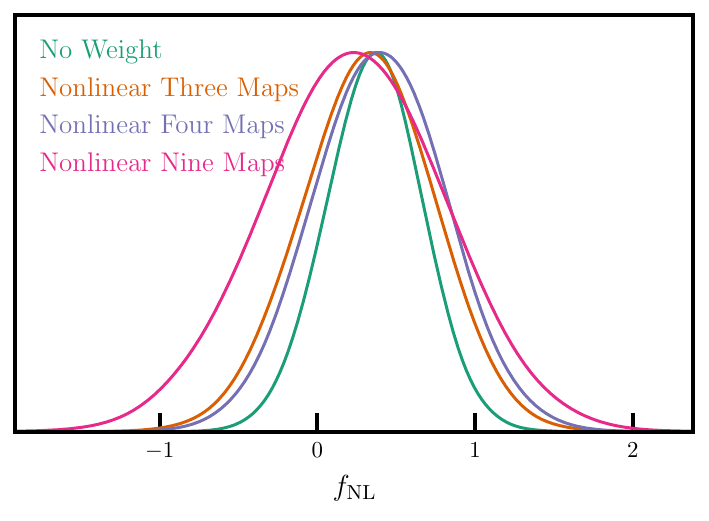}
\includegraphics[width=0.45\textwidth]{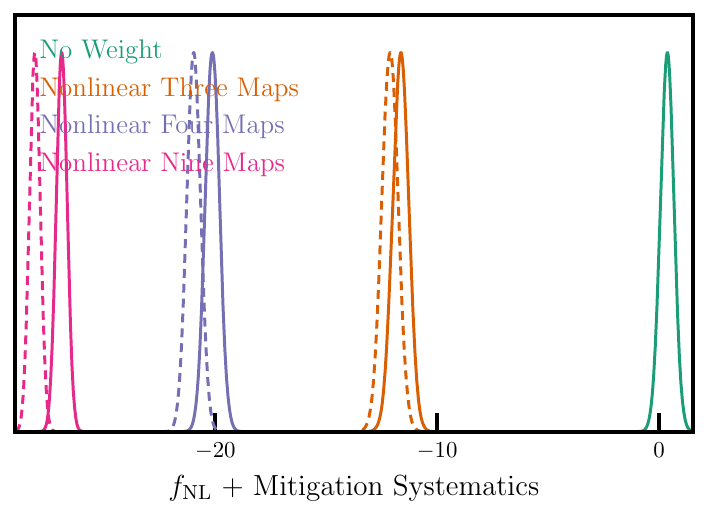}
\includegraphics[width=0.45\textwidth]{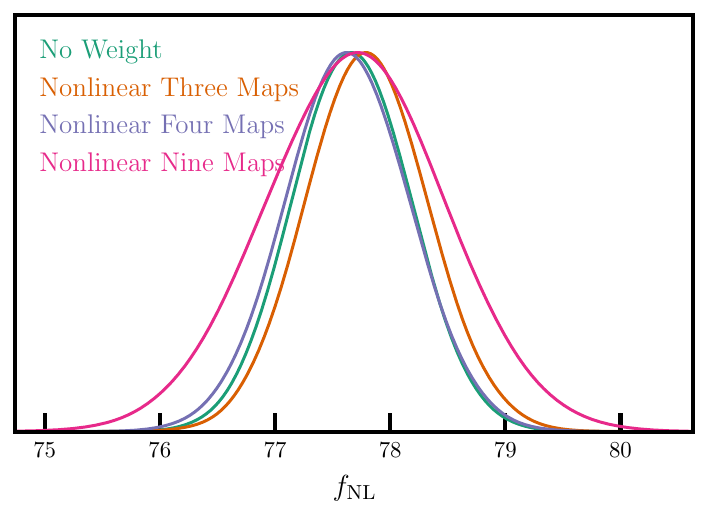}
\includegraphics[width=0.45\textwidth]{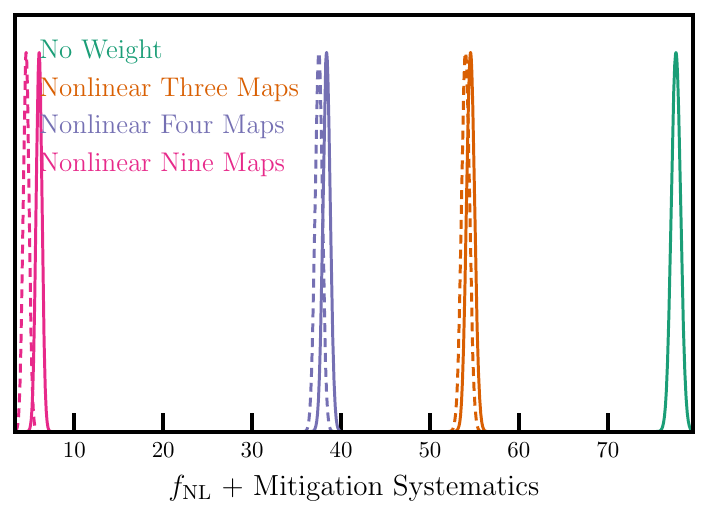}
\caption{Probability distributions of $\fnl$ from the mean power spectrum of the $\fnl=0$ (top) and $\fnl=76.9$ (bottom) mocks before and after mitigation with the non-linear methods using three, four, and nine maps. The dashed (solid) curves show the distributions for the contaminated (clean) mocks. Left: The posteriors are adjusted to account for the over-correction effect. Right: The posteriors are subject to the over-correction effect, and thus the scaling of $\fnl$ values is biased due to mitigation.}\label{fig:contmcmc}
\end{figure*}

\begin{table*}
    \caption{The best-fitting and marginalized estimates for $\fnl$ from fitting the mean power spectrum of the mocks before and after corrections using the non-linear approach with various combinations of the imaging systematic maps. The covariance is scaled to represent the error on the mean power spectrum. The estimates are not accounted for over-correction, and therefore are subject to mitigation systematics.}   
    \label{tab:contmocksmcmc}  
   \centerline{%
    \begin{tabular}{lllllll}
    \hline
    \hline
   &  & 	   & & $\fnl$ + Mitigation Systematics & & \\
   \cmidrule(r{.7cm}){3-6}    
Mock / $\fnl$ & Method & Best fit  & Mean & $ 68\%$ CL & $ 95\%$ CL & $\chi^{2}$ (dof $=34$) \\
    \hline
Clean $0$ & No Weight                   & $  0.36$& $  0.36$& $  0.06<\fnl<  0.65$& $ -0.23<\fnl<  0.94$ &   35.7\\
Clean $0$ & Three Maps                  & $-11.64$& $-11.65$& $-12.00<\fnl<-11.30$& $-12.34<\fnl<-10.97$ &   86.8\\
Clean $0$ & Four Maps                   & $-20.14$& $-20.13$& $-20.44<\fnl<-19.82$& $-20.74<\fnl<-19.52$ &  472.8\\
Clean $0$ & Nine Maps                   & $-26.91$& $-26.92$& $-27.16<\fnl<-26.68$& $-27.39<\fnl<-26.46$ & 5481.0\\
Contaminated $0$ & Three Maps           & $-12.12$& $-12.13$& $-12.48<\fnl<-11.78$& $-12.83<\fnl<-11.44$ &   94.0\\
Contaminated $0$ & Four Maps            & $-20.97$& $-20.98$& $-21.28<\fnl<-20.67$& $-21.58<\fnl<-20.37$ &  556.3\\
Contaminated $0$ & Nine Maps            & $-28.13$& $-28.13$& $-28.36<\fnl<-27.90$& $-28.59<\fnl<-27.67$ & 6760.5\\
\hline
Clean $76.9$ & No Weight               & $ 77.67$& $ 77.67$& $ 77.17<\fnl< 78.16$& $ 76.71<\fnl< 78.64$ &   38.8\\
Clean $76.9$ & Three Maps              & $ 54.57$& $ 54.57$& $ 54.14<\fnl< 55.01$& $ 53.72<\fnl< 55.45$ &  603.5\\
Clean $76.9$ & Four Maps               & $ 38.38$& $ 38.38$& $ 37.99<\fnl< 38.78$& $ 37.60<\fnl< 39.16$ &  537.0\\
Clean $76.9$ & Nine Maps               & $  6.04$& $  6.04$& $  5.72<\fnl<  6.36$& $  5.41<\fnl<  6.67$ &  694.0\\
Contaminated $76.9$ & Three Maps       & $ 54.01$& $ 54.00$& $ 53.57<\fnl< 54.44$& $ 53.15<\fnl< 54.86$ &  588.0\\
Contaminated $76.9$ & Four Maps        & $ 37.48$& $ 37.49$& $ 37.09<\fnl< 37.88$& $ 36.70<\fnl< 38.27$ &  510.7\\
Contaminated $76.9$ & Nine Maps        & $  4.59$& $  4.58$& $  4.26<\fnl<  4.90$& $  3.95<\fnl<  5.22$ &  649.7\\
\hline
    \end{tabular}
    }
\end{table*}
Using the calibration parameters presented in \S \ref{ssec:calibration}, we account for the shift in the $\fnl$ constraints caused by the imaging systematic mitigation. We show the marginalized probability distributions on $\fnl$ before and after accounting for the over-correction in the right and left panels of Figure \ref{fig:contmcmc}.

